\input amstex
\input epsf
\magnification=\magstep1 
\baselineskip=13pt
\documentstyle{amsppt}
\vsize=8.7truein \CenteredTagsOnSplits \NoRunningHeads
\def\EE{\bold {E\thinspace}}
\def\ii{\bold{i}}
\def\vl{\operatorname{vol}}
\topmatter
 
\title Computing Gaussian and exponential integrals in ${\Bbb R}^n$ \endtitle 
\author Alexander Barvinok  \endauthor
\address Department of Mathematics, University of Michigan, Ann Arbor,
MI 48109-1043, USA \endaddress
\email barvinok$\@$umich.edu \endemail
\date August 12, 2026 \enddate
\thanks  
\endthanks 
\abstract We consider expectations of the type $\EE \exp \left\{\sum_{i=1}^m \phi_i \right\}$, where $\phi_i: {\Bbb R}^n \longrightarrow {\Bbb C}$ are functions, 
each depending on a few coordinates of a point in ${\Bbb R}^n$, and the expectation is taken with respect to the standard Gaussian or symmetric exponential probability measures. We prove sufficient conditions, in terms of the Lipschitz constants of $\phi_i$ and the combinatorics of their dependencies, for the integral to be non-zero, and, consequently, to be amenable to a computationally efficient approximation. We discuss applications to computing volumes of bodies and statistics on integer points in polyhedra in ${\Bbb R}^n$.
\endabstract
\endtopmatter
\document

\head 1. Introduction and main results \endhead 

We construct efficient deterministic algorithms to approximate expectations $\EE e^f$ for complex-valued functions $f: {\Bbb R}^n \longrightarrow {\Bbb C}$, with respect to the standard Gaussian and symmetric exponential probability measures in ${\Bbb R}^n$. The functions $f$ are represented as sums
$f=\phi_1 + \ldots + \phi_m$, where the dimension $n$ of the ambient space and the number $m$ of terms can be large, but where each $\phi_i$ depends on a few coordinates of 
a point $x=\left(\xi_1, \ldots, \xi_n\right)$, $x \in {\Bbb R}^n$,  and we control the dependencies among $\phi_i$ and the Lipschitz constants of $\phi_i$. We apply the results to compute statistics for integer points in higher-dimensional polyhedra and to approximate volumes of some higher-dimensional bodies, including some non-convex.

\subhead (1.1) The setup \endsubhead Let $\mu=\mu_1 \times \cdots \times \mu_n$ be the product probability measure in 
${\Bbb R}^n ={\Bbb R} \oplus \cdots \oplus {\Bbb R}$ and let $\phi_1, \ldots, \phi_m: {\Bbb R}^n \longrightarrow {\Bbb C}$ be complex-valued random variables. We are interested in efficient computation (approximation) of the expectation 
$$\EE \exp\left\{ \sum_{i=1}^m \phi_i \right\}. \tag1.1.1$$
Of course, as stated, the integral (1.1.1) is way too general. We will assume that each function $\phi_i$ depends only on a few coordinates of a point 
$x=\left(\xi_1, \ldots, \xi_n\right)$ and impose some restrictions on the dependencies between the functions. We will also control the Lipschitz constants of $\phi_i$.

Integrals of the type (1.1.1) are ubiquitous in statistical physics and quantum field theory, see, for example, \cite{Br86}, \cite{GJ87} and \cite{FV18}. Here we are interested in the computational complexity issues, as well as some discrete geometry applications. 

The problem of approximating (1.1.1) is closely related to the problem of deciding when (1.1.1) is not 0. We say that complex numbers $z_1 \ne 0$ and $z_2 \ne 0$ approximate each other within a relative error of $\epsilon >0$ if we can write $z_1=e^{w_1}$ and $z_2=e^{w_2}$ for some numbers $w_1, w_2 \in {\Bbb C}$ such that $|w_1-w_2| \leq \epsilon$. It is immediately clear that having (1.1.1) equal 0 creates a difficulty: in any meaningful way, a relative approximation 
of 0 can only be 0.

As it turns out, having (1.1.1) guaranteed to be non-zero opens a way to efficiently approximate the integral. This idea has been around for some time by now, see \cite{Ba16} and \cite{PR17}. Here we briefly sketch how it is done, providing more detail in Section 2.

We introduce a parameter $z \in {\Bbb C}$ and, given functions $\phi_1, \ldots, \phi_m: {\Bbb R}^n \longrightarrow {\Bbb C}$, consider the expectation 
$$g(z)=\EE \exp\left\{ z \sum_{i=1}^n \phi_i \right\} \tag1.1.2$$
as a function of $z$. Let us fix a $\rho >1$. Suppose that for some functions $\phi_i$ and some $M \geq 3$, we have 
$$0 \ < \ |g(z)| \ \leq \ M \quad \text{provided} \quad |z| \leq \rho. \tag1.1.3$$

 It turns out then that one can approximate (1.1.1) within relative error $0 < \epsilon < 1$ in polynomial time from the moments 
$$\split &\EE \left( \sum_{i=1}^m \phi_i\right)^l =\sum_{1 \leq i_1, \ldots, i_l \leq m} \EE \left( \phi_{i_1} \cdots \phi_{i_l} \right)\\ &\qquad \text{for} \quad l\leq s = 
O_{\rho}\left( \ln \ln M - \ln \epsilon \right), \endsplit \tag1.1.4$$
where the implied constant in the ``$O$" notation depends only on $\rho$. In some cases,  (1.1.3) holds with $M=\exp\left\{ (mn)^{O(1)}\right\}$, which results in the bound $$s=O_{\rho}\left(\ln m + \ln n -\ln \epsilon\right)$$ in (1.1.4). We note that while the upper bound in (1.1.3) is usually quite straightforward, it is the lower bound, that is, showing that $g(z) \ne 0$ for all $|z| \leq \rho$, that requires work.

 Assuming that each $\phi_i$ depends on $r_i=O(1)$ coordinates, computing each of the not more than $m^{s+1}$ expectations $\EE\left( \phi_{i_1} \cdots \phi_{i_l}\right)$ in (1.1.4)
reduces to integration in a coordinate subspace of dimension $r=r_{i_1} + \ldots + r_{i_l}=O(l)$. In most applications, integration in a $l$-dimensional space can be done in  
$l^{O(l)}$ time, which in turn produces a quasi-polynomial algorithm of $(m+n)^{O_{\rho}(\ln ((m+n)/\epsilon))}$ complexity to approximate (1.1.1).

In fact, to approximate (1.1.1), one can replace (1.1.3) by a weaker condition: it suffices to have the inequality satisfied for all $z$ in some fixed connected open set ${\Bbb U} \subset {\Bbb C}$ containing $0$ and $1$, and not necessarily in the disc $|z| \leq \rho$, cf. Section 2.2 of \cite{Ba16}.

Zeros of $g(z)$ in (1.1.2) are of a considerable interest to statistical physics, as they correspond to phase transitions, with $z$ playing the role of the inverse temperature, see \cite{FV18}. One can informally say that if the system stays sufficiently far away from a phase transition, then the partition function can be efficiently approximated. In probability, the absence of zeros is related to the Central Limit Theorem type behavior of sequences \cite{MS26}.

In this paper, we consider two special cases. In the first case,  $\mu$ is the standard Gaussian measure with density
$${1 \over (2\pi)^{n/2}} \exp\left\{ -{1 \over 2} \sum_{i=1}^n \xi_i^2 \right\} \quad \text{where} \quad x=\left(\xi_1, \ldots, \xi_n\right), \tag1.1.5$$
and we control the Lipschitz constants of $\phi_i$ in the $\ell^2$ norm. In the second case, 
$\mu$ is the symmetric exponential measure with density 
$${1 \over 2^n} \exp\left\{ -\sum_{j=1}^n |\xi_j| \right\} \quad \text{for} \quad x=\left(\xi_1, \ldots, \xi_n\right), \tag1.1.6$$
and we control the Lipschitz constants of $\phi_i$ in the $\ell^1$ norm.
In \cite{Ba26}, the author considered the integral (1.1.1) in the case of a general product probability measure $\mu$ in the product space $\Omega=\Omega_1 \times \cdots \times \Omega_n$, with controlled Lipschitz constants of $\phi_i$ in the Hamming metric of $\Omega$.

For computational purposes, it is convenient, and is not really restrictive, to assume that 
$$\phi_i(0)=0 \quad \text{for} \quad i=1, \ldots, m, \tag1.1.7$$
since replacing $\phi_i$ by $\phi_i-\phi_i(0)$ results in multiplying (1.1.1) by $e^{-\phi_i(0)}$. The condition (1.1.7) also allows us to avoid dealing with potentially very large numbers, of an exponential bit size.

To control the dependencies among $\phi_i$, we introduce some formal definitions.
We say that a function $\phi: {\Bbb R}^n \longrightarrow {\Bbb C}$ {\it depends on the coordinates} $\left\{ \xi_j:\ j \in J_{\phi}\right\}$ provided
$$\phi\left(\xi_1', \ldots, \xi_n'\right)=\phi\left(\xi_1'', \ldots, \xi_n''\right) \quad \text{whenever} \quad \xi_j' = \xi_j'' \quad \text{for all} \quad j \in J_{\phi},$$
and $J_{\phi}$ is the minimal set under inclusion with that property. 
We say that $\phi$ {\it depends on at most $r$ coordinates} if $|J_{\phi}| \leq r$.
We say that functions $\phi, \psi: {\Bbb R}^n \longrightarrow {\Bbb C}$ {\it share a coordinate} if $J_{\phi} \cap J_{\psi} \ne \emptyset$. 

A popular approach to analyze the integrals (1.1.1) and (1.1.2) is via the {\it cluster expansion}, see \cite{Br86}. Here we pursue a different, inductive approach that seems to produce stronger results in terms of the combinatorics of dependencies and also in terms of the required analytic properties of functions $\phi_i$. Thus, unlike in the case of the cluster expansion approach \cite{Br86} for the Gaussian measure, we do not require to bound higher derivatives of $\phi_i$ or even to assume that the functions are smooth. The proofs for the Gaussian and exponential measures presented in this paper are quite similar and can be extended to other measures with sufficiently strong concentration properties and a suitable logarithmic Sobolev inequality, cf. Chapter 5 of \cite{Le01}.

Our first result deals with the Gaussian measure with density (1.1.5).
We consider the standard $\ell^2$ norm in ${\Bbb R}^n$:
$$\|x\|_2=\left(\xi_1^2 + \ldots + \xi_n^2\right)^{1/2} \quad \text{where} \quad x=\left(\xi_1, \ldots, \xi_n\right).$$
For $L >0$, a function $\phi: {\Bbb R}^n \longrightarrow {\Bbb C}$ is $L$-{\it Lipschitz} in the $\ell^2$ norm, provided 
$$| \phi(x) - \phi(y)| \ \leq \ L\|x-y\|_2.$$
We prove the following result.

\proclaim{(1.2) Theorem} Let $\phi_1, \ldots, \phi_m: {\Bbb R}^n \longrightarrow {\Bbb C}$ be functions such that $\phi_i$ is $L_i$-Lipschitz in the $\ell^2$ norm
for some $L_i \geq 0$ and $i=1, \ldots, m$. For $k=1, \ldots, m$, let $I_k \subset \{1, \ldots, m\}$ be the set of indices $i$ such that 
$\phi_i$ shares a coordinate with $\phi_k$ (in particular, $k \in I_k$ unless $\phi_k$ is constant and $I_k=\emptyset$).
 Suppose that 
\roster
\item For some $c \geq 1$ and all $j=1, \ldots, n$, at most $c$ functions $\phi_i$ depend on the coordinate $\xi_j$ and
\item We have 
$$\sum_{i \in I_k} L_i^2 \ \leq \ {1 \over 64 c} \quad \text{for} \quad k=1, \ldots, m.$$
\endroster
Then
$$\EE \exp\left\{ \sum_{i=1}^m \phi_i \right\} \ne 0,$$
where the expectation is taken with respect to the standard Gaussian probability measure with density (1.1.5).
\endproclaim 

We are interested in the situations where the parameter $c$ is small (fixed in advance), whereas $m$ and $n$ are allowed to grow.

As the Gaussian measure with density (1.1.5) is invariant under orthogonal transformations of ${\Bbb R}^n$, one would naturally inquire if Theorem 1.2 can be restated in an invariant language, without a reference to a specific coordinate system. This is indeed the case. Let $V$ be a Euclidean space endowed withe the standard Gaussian probability measure and let $P_1, \ldots, P_m: V \longrightarrow V$ be pairwise commuting orthogonal projections onto subspaces of $V$, so that $P_i^2 = P_i$ for $i=1, \ldots, m$ and $P_i P_j =P_j P_i$ for all $1 \leq i, j \leq m$. Suppose that $c \geq 1$ is the largest integer for which there exists a product of $c$ distinct operators $P_i$ that is not $0$. Let $\phi_1, \ldots, \phi_m: V \longrightarrow {\Bbb C}$ be functions such that $\phi_i$ is $L_i$-Lipschitz for $i=1, \ldots, m$, and suppose that 
$$\sum_{i:\ P_i P_k \ne 0} L_i^2 \ \leq \ {1 \over 64 c} \quad \text{for} \quad k =1, \ldots, m.$$
Then 
$$\EE \exp\left\{ \sum_{i=1}^m \phi_i(P_i(x)) \right\} \ne 0.$$

Let functions $\phi_1, \ldots, \phi_m$, their Lipschitz constants $L_1, \ldots, L_m$, and the parameter $c$  be as in Theorem 1.2. Let us fix a $\rho >1$ and suppose that 
Part (2) of Theorem 1.2 is satisfied with the slack $\rho$, that is
$$\sum_{i \in I_k} L_i^2 \ \leq \ {1 \over 64 \rho c} \quad \text{for} \quad k=1, \ldots, m.$$
We also assume that (1.1.7) holds.
 As we argue in Section 2,
 the expectation 
$$\EE \exp\left\{  \sum_{i=1}^m  \phi_i \right\}$$
can be approximated within relative error $0< \epsilon < 1$ in polynomial time from the moments (1.1.4) with $s=O_{\rho} \bigl(\ln (m+n) -\ln \epsilon \bigr)$,
where the implied constant in the ``$O$'' notation depends only on $\rho$, see Theorem 2.5 for the exact statement.

A weaker version of Theorem 1.2 is obtained in \cite{Ba26}: there the conclusion $\EE \exp\left\{ \sum_{i=1}^m  \phi_i\right\} \ne 0$ is proven to hold,
assuming that the functions $\phi_i$ are $L_i$-Lipschitz in the $\ell^1$ norm of ${\Bbb R}^n$, each function $\phi_i$ depends on at most $r \geq 2$ coordinates, and that
$$\sum_{i \in I_j} L_i \ \leq \ {1 \over 6 \sqrt{r}} \quad \text{for} \quad j=1, \ldots, n,$$
where $I_j$ consists of all indices $i$ such that $\phi_i$ depends on the $j$-th coordinate $\xi_j$. In particular, the version of \cite{Ba26} is not invariant under orthogonal transformations of ${\Bbb R}^n$.

We prove Theorem 1.2 in Section 5, and in Section 3 we describe some applications to integer point counting in polyhedra. 

Our second result deals with the symmetric exponential measure with density (1.1.6).
We consider the $\ell^{1}$ norm in ${\Bbb R}^n$:
$$\|x\|_1=|\xi_1| + \ldots + |\xi_n| \quad \text{where} \quad x=\left(\xi_1, \ldots, \xi_n\right).$$
For $L > 0$, a function $\phi: {\Bbb R}^n \longrightarrow {\Bbb C}$ is {\it $L$-Lipschitz} in the $\ell^1$ norm, provided
$$|\phi(x)-\phi(y)| \ \leq \ L \|x-y\|_1.$$
Our main result is as follows.

\proclaim{(1.3) Theorem} Let $\phi_1, \ldots, \phi_m: {\Bbb R}^n \longrightarrow {\Bbb C}$ be functions such that $\phi_i$ is $L_i$-Lipschitz in the $\ell^1$ norm 
for some $L_i \geq 0$ and $i=1, \ldots, m$. For $j=1, \ldots, n$, let 
$I_j \subset \{1, \ldots, m\}$ be the set of indices $i$ such that $\phi_i$ depends on the coordinate $\xi_j$.
Suppose that 
\roster
\item For some $r \geq 9$, each function $\phi_i$ depends on at most $r$ coordinates and 
 \item We have 
$$\sum_{i \in I_j} L_i \ \leq \ {1 \over 24 \sqrt{r}} \quad \text{for} \quad j=1, \ldots, n.$$
\endroster
Then $$ \EE \exp\left\{  \sum_{i=1}^m  \phi_i \right\}\ \ne \ 0,$$
where the expectation is taken with respect to the symmetric exponential probability measure with density (1.1.6).
\endproclaim

Again, we are interested in the situations where $r$ is small (fixed in advance), whereas $m$ and $n$ are allowed to grow.

 Let functions $\phi_i$, their Lipschitz constants $L_i$ and the parameters $r$ be as in Theorem 1.3. Let us fix a $\rho >1$ and suppose that Part (2) of Theorem 1.3 holds with the slack $\rho$, that is 
 $$\sum_{i \in I_j} L_i \ \leq \ {1 \over 24 \rho \sqrt{r}} \quad \text{for} \quad j=1, \ldots, n.$$
 We also assume that (1.1.7) holds.
 Then 
 $$\EE \exp\left\{  \sum_{i=1}^m \phi_i \right\}$$
can be approximated within relative error $0 < \epsilon < 1$ in polynomial time from the moments (1.1.4) with $s=O_{\rho}\bigl(\ln n - \ln \epsilon \bigr)$, see Theorem 2.7 for the exact statement.

We prove Theorem 1.3 in Section 6, and in Section 4 we describe some applications to computing volumes.

\subhead (1.4) Notation \endsubhead We denote by $\ii$ the imaginary unit, so $\ii^2=-1$. For a complex number $z=a+\ii b$, we denote by $\Re$ the real part, and by $\Im$ the imaginary part: $\Re\thinspace z = a$, $\Im\thinspace z=b$.

\head 2. Computing approximations \endhead 

Here we discuss in some detail how Theorems 1.2 and 1.3 lead to approximation algorithms. 
For $\rho >1$, let 
$${\Bbb D}_{\rho}=\left\{ z:\ |z| \leq  \rho \right\}$$ be 
the closed disc of radius $\rho$ in the complex plane, centered at $0$.
\subhead (2.1) Approximating within an additive error \endsubhead  
 Let $f: {\Bbb D}_{\rho} \longrightarrow {\Bbb C}$ be a holomorphic function such that
$$|f(z)| \ \leq \ M \quad \text{for all} \quad z \in {\Bbb D}_{\rho}$$
and some $M > 1$.
For an integer $s >0$, let 
$$T_s(f; z)=f(0) + \sum_{l=1}^s {f^{(l)}(0)\over l!} z^l$$
be the Taylor polynomial of $f$ of degree $s$ computed at $0$. 
The Cauchy bound,
$$\left| {f^{l}(0) \over l!} \right| \ \leq \ {M \over \rho^l},$$
see, for example, Section III.7 of \cite{La99}, implies that 
$$\left| f(1)-T_s(f; 1)\right| = \left| \sum_{l=s+1}^{\infty} {f^{(l)}(0) \over l!}\right| \ \leq \ M \sum_{l=s+1}^{\infty} \rho^{-l} = 
{M \over (\rho-1) \rho^s}. \tag2.1.1$$
It follows from (2.1.1) that to approximate $f(1)$ by $T_s(f; 1)$ within an additive error of $0 < \epsilon < 1$, it suffices to choose 
$s=O_{\rho}\bigl(\ln M - \ln \epsilon \bigr)$, where the implied constant in the ``$O$" notation depends on $\rho$ only.

\subhead (2.2) Approximating within a relative error \endsubhead
Suppose now that $g: {\Bbb D}_{\rho} \longrightarrow {\Bbb C}$ is a holomorphic function such that 
$$g(0)=1 \quad \text{and} \quad 0 \ < \ |g(z)| \ \leq \ M \quad \text{for all} \quad z \in {\Bbb D}_{\rho}\tag2.2.1$$
and some $M \geq 3 $. Our goal is to approximate $g(1)$ within relative error $0 < \epsilon <1$.

Since $g(z) \ne 0$ for all $z \in {\Bbb D}_{\rho}$, we can choose a continuous branch of 
$$f(z)=\ln g(z),$$
which we choose so that $f(0)=\ln 1 =0$. 

From (2.2.1) we have 
$$\Re\thinspace f(z) \ \leq \ \ln M \quad \text{for all} \quad z \in {\Bbb D}_{\rho}. \tag2.2.2$$
The Borel - Carath\'eodory Theorem (see, for example, Section XII.3 of \cite{La99}) asserts that 
$$\max_{|z| \leq r} |f(z)| \ \leq \ {2r \over \rho - r} \sup_{|z| \leq \rho} \Re \thinspace f(z) + {\rho + r \over \rho-r} |f(0)| \quad \text{for} \quad 0 < r < \rho,$$
which, together with (2.2.2), implies that 
$$|f(z)| \ \leq \ {2\rho+2 \over \rho-1} \ln M  \quad \text{for all} \quad z \in {\Bbb D}_{{\rho+1 \over 2}}. \tag2.2.3$$

As in Section 2.1, we approximate $f(1)$ within additive error $0 < \epsilon < 1$ by the value $T_s(f; 1)$ of the Taylor polynomial of $f$ and hence approximate 
$g(1)$ by $e^{f(1)}$ within relative error $\epsilon$. It follows from (2.1.1) and (2.2.3) that we can choose 
$$s=O_{\rho} \left( \ln \ln M - \ln \epsilon\right). \tag2.2.4$$

Moreover, one can compute the derivatives $f^{(l)}(0)$ for $l=1, \ldots, s$ from the derivatives $g^{(l)}(0)$ for $l=1, \ldots, s$ in $O(s^2)$ time by solving a $s \times s$ non-degenerate triangular system of linear equations, so that the Taylor polynomial $T_s(f; z)$ of $f=\ln g$ can be computed in $O(s^2)$ time from the Taylor polynomial $T_s(g; z)$ of $g$. Indeed, since $f'(z) = g'(z)/g(z)$, we have $g'(z)=f'(z)g(z)$ and hence 
$$g^{(l)}(0)=\sum_{k=0}^{l-1} {l-1 \choose k} f^{(l-k)}(0) g^{(k)}(0),$$
see Section 2.2 of \cite{Ba16} for detail.

We also remark that (2.2.1) can be replaced by a weaker condition that the inequalities hold for all $z$ in some fixed connected open set ${\Bbb U} \subset {\Bbb C}$, containing 0 and 1. We can reduce this general case to that of the disc ${\Bbb D}_{\rho}$, by replacing $g(z)$ with the composition 
$g(w(z))$, where $w: {\Bbb D}_{\rho} \longrightarrow {\Bbb U}$ is a holomorphic map, such that $w(0)=0$ and $w(1)=1$, see Section 2.2 of \cite{Ba16}.

\subhead (2.3) Approximating expectations \endsubhead 
Given functions $\phi_1, \ldots, \phi_m: {\Bbb R}^n \longrightarrow {\Bbb C}$, our goal is to approximate (1.1.1)
within relative error $0 < \epsilon <1$. We assume that (1.1.7) holds.
We introduce a complex parameter $z$ and consider 
$$g(z)=\EE \exp\left\{ z \sum_{i=1}^m \phi_i \right\}. \tag2.3.1$$
We assume that for some $\rho > 1$ the integral (2.3.1) converges absolutely and uniformly on the disc ${\Bbb D}_{\rho}= \{z:\ |z| \leq \rho \}$ in the complex plane, and, moreover, that for some $M \geq 3$ we have 
$$0 \ < \ |g(z)| \ \leq \ M \quad \text{for all} \quad z \in {\Bbb D}_{\rho}. \tag2.3.2$$
We note that $g(0)=1$ and that the expectation (1.1.1) is $g(1)$.
It follows from Section 2.2, see (2.2.4) in particular, that one can approximate $g(1)$ within relative error $0 < \epsilon < 1$ in polynomial time 
from the moments
$$\split &g^{(l)}(0)= \EE \left( \sum_{i=1}^m \phi_i \right)^l = \sum_{1 \leq i_1, \ldots, i_l \leq m} \EE\left( \phi_{i_1} \cdots \phi_{i_l}\right) \\ &\qquad  \text{for} \quad l \leq s=O_{\rho}\left( \ln \ln M - \ln \epsilon \right), \endsplit
\tag2.3.3$$
where the implied constant in the ``$O$" notation depends on $\rho$ only.

In the next two sections, we obtain bounds for $M$ in the case of the standard Gaussian (Section 2.4 and Theorem 2.5) and symmetric exponential (Section 2.6 and Theorem 2.7) measures, in the context of Theorem 1.2 and Theorem 1.3 respectively.

\subhead (2.4)  Gaussian measure \endsubhead Let $\phi_1, \ldots, \phi_m: {\Bbb R}^n \longrightarrow {\Bbb C}$ be functions 
 such that $\phi_i$ is $L_i$-Lipschitz in the $\ell^2$ norm and $\phi_i(0)=0$ for $i=1, \ldots, m$. We assume that Part (2) of Theorem 1.2 holds with the slack 
 $1< \rho < 2$, that is 
$$\sum_{i \in I_k} L_i^2 \ \leq \ {1 \over 64\rho c} \quad \text{for} \quad k=1, \ldots, m.$$
We define $g: {\Bbb D}_{\rho} \longrightarrow {\Bbb C}$ by (2.3.1) and our immediate goal is to bound $M$ in (4.3.2).
 Let $x', x'' \in {\Bbb R}^n$, $x'=\left(\xi_1', \ldots, \xi_n\right)$ 
and $x''=\left(\xi_1'', \ldots, \xi_n''\right)$. Suppose that $\phi_i$ depends on the coordinates $\xi_j$ with $j \in J_i$ and let $x'_i, x''_i \in {\Bbb R}^{J_i}$ be the vectors obtained from $x'$, respectively $x''$, by picking up the coordinates $\xi_j$ with $j \in J_i$. 
Then for any $z \in {\Bbb D}_{\rho}$, applying the Cauchy - Schwarz inequality, we obtain
$$\split &\left| z \sum_{i=1}^m  \phi_i(x') - z \sum_{i=1}^m\phi_i(x'') \right| \ \leq \ \rho \sum_{i=1}^m L_i \| x_i' - x_i''\|_2 \\ &\qquad \leq \ 
\rho \left(\sum_{i=1}^m L_i^2 \right)^{1/2} \left( \sum_{i=1}^m \|x_i' - x_i''\|_2^2 \right)^{1/2}. \endsplit $$
We bound 
$$\sum_{i=1}^m L_i^2 \ \leq \ \sum_{k=1}^m \sum_{i \in I_k} L_i^2  \ \leq \ {m \over 64 \rho c}.$$ 
Since for each coordinate $\xi_j$ there are at most $c$ sets $J_i$ with $j \in J_i$, we have 
$$\sum_{i=1}^m \|x_i' - x_i''\|_2^2 \ \leq \ c \|x'-x''\|_2^2.$$ Summarizing, for any $z \in {\Bbb D}_{\rho}$, the function $\psi: {\Bbb R}^n \longrightarrow {\Bbb C}$,
$$\psi(x)=z \sum_{i=1}^m \phi_i(x)$$ is
$L$-Lipschitz with in the $\ell^2$ norm with
$$L=\rho \left({m \over 64 \rho c}\right)^{1/2} c^{1/2} = {1 \over 8} \sqrt{\rho m} \ \leq \ \sqrt{m}.$$

Since $\psi(0)=0$, we have 
$$|\psi(x)| \ \leq \ \sqrt{m} \|x\|_2 \quad \text{for all} \quad x \in {\Bbb R}^n.$$
Then
$$\aligned &\left| \EE e^{\psi}\right|  \ \leq \ \EE e^{\sqrt{m} \|x\|_2} = {1 \over (2\pi)^{n/2}} \int_{{\Bbb R}^n} \exp\left\{\sqrt{ m} \|x\|_2 - {1 \over 2} \|x\|_2^2 \right\} \ dx \\
&\qquad ={1 \over (2\pi)^{n/2}} \int_{{\Bbb R}^n} \exp\left\{ \sqrt{m} \|x\|_2 - {1 \over 4} \|x\|_2^2 \right\} e^{-\|x\|_2^2/4} \ dx \\
&\qquad \ \leq \ {e^{m} \over (2\pi)^{n/2}} \int_{{\Bbb R}^n} e^{-\|x\|_2^2/4} \ dx =e^{m} 2^{n/2}. \endaligned \tag2.4.1$$
From (2.4.1) and Theorem 1.2, it follows that for the function $g(z)$ defined by (2.3.1), we can choose $M=e^m 2^{n/2}$ and hence in (2.3.3) we have 
$$s=O_{\rho}\left(\ln(m+n)-\ln \epsilon \right).$$

We summarize the results of Sections 2.1-2.4. 

\proclaim{(2.5) Theorem} Fix a $\rho > 1$. Let $\phi_i: {\Bbb R}^n \longrightarrow {\Bbb C}$ be functions such that $\phi_i$ is $L_i$-Lipschitz in the $\ell^2$ norm 
for some $L_i \geq 0$ and $i=1, \ldots, m$. For $k=1, \ldots, m$, let $I_k \subset \{1, \ldots, m\}$ be the set of indices $i$ such that $\phi_i$ shares a coordinate with $\phi_k$.
Suppose that 
\roster
\item For some $c \geq 1$ and all $j=1, \ldots, n$, at most $c$ functions $\phi_i$ depend on the coordinate $\xi_j$;
\item We have 
$$\sum_{i \in I_k} L_i^2 \ \leq \ {1 \over 64 \rho c} \quad \text{for} \quad k=1, \ldots, m;$$
\item We have 
$$\phi_i(0)=0 \quad \text{for} \quad i=1, \ldots, m.$$
\endroster
Then, for any $0 < \epsilon < 1$, the expectation
$$\EE \exp\left\{ \sum_{i=1}^m \phi_i \right\}$$ 
with respect to the standard Gaussian measure with density (1.1.5)
can be approximated within relative error $\epsilon$  in polynomial time from the mixed moments 
$$\EE \left(\phi_{i_1} \cdots \phi_{i_l}\right) \quad \text{where} \quad 1 \leq i_1, \ldots, i_l \leq m, \quad l \leq  s = O_{\rho} \left(\ln (m+n) - \ln \epsilon\right),$$
where the implied constant in the ``$O$" notation depends on $\rho$ only.
\endproclaim
{\hfill \hfill \hfill} \qed 

\subhead (2.6) Exponential measure \endsubhead Let $\phi_1, \ldots, \phi_m: {\Bbb R}^n \longrightarrow {\Bbb C}$ be functions such that $\phi_i$ is $L_i$-Lipschitz in the $\ell^1$ norm and that $\phi_i(0)=0$ for $i=1, \ldots, m$. We assume that constraint of Theorem 1.3 holds with some slack $\rho > 1$, that is, 
$$\sum_{i \in I_j} L_i \ \leq \ {1 \over 24 \rho \sqrt{r}} \quad \text{for} \quad j=1, \ldots, n.$$
We define $g: {\Bbb D}_{\rho} \longrightarrow {\Bbb C}$ by (2.3.1) and our immediate goal is to bound $M$ in (4.3.2). Suppose that $\phi_i$ depends on the coordinates $\xi_j$ with $j \in J_i$ and let $x'_i, x''_i \in {\Bbb R}^{J_i}$ be the vectors obtained from $x'$, respectively $x''$, by picking up the coordinates $\xi_j$ with $j \in J_i$. Then for $z \in {\Bbb D}_{\rho}$, we have
$$\split &\left| z \sum_{i=1}^m \phi_i(x') - z \sum_{i=1}^m \phi_i(x'')\right| \ \leq \ \rho \sum_{i=1}^m L_i \|x'_i - x''_i\|_1 = \rho \sum_{j =1}^n  \left( \sum_{i:\ i \in I_j} L_i\right) 
|\xi_j'-\xi_j''| \\ &\qquad \leq \ \rho { 1 \over 24 \rho \sqrt{r}} \| x'-x''\|_1= {1 \over 24 \sqrt{r}} \left\| x'- x'' \right\|_1 \ \leq \ {1 \over 2} \|x'-x''\|_1. \endsplit $$
Summarizing, for any $z \in {\Bbb D}_{\rho}$, the function $\psi: {\Bbb R}^n \longrightarrow {\Bbb C}$ defined by
$$\psi(x)=z \sum_{i=1}^m \phi_i(x) \quad \text{for} \quad x \in {\Bbb R}^n$$
is ${1 \over 2}$-Lipschitz in the $\ell^1$ norm of ${\Bbb R}^n$. Since $\psi(0)=0$, we have 
$$|\psi(x)| \ \leq \ {1 \over 2} \|x\|_1 \quad \text{for all} \quad x \in {\Bbb R}^n.$$
Then
$$\left| \EE e^{\psi} \right| \ \leq \ \EE \left| e^{\psi(x)}\right|  \ \leq \ {1 \over 2^n} \int_{{\Bbb R}^n} \exp\left\{ -{1 \over 2} \sum_{j=1}^n |\xi_j| \right\} \ d x=2^n. \tag2.6.1$$
It follows from (2.6.1) and Theorem 1.3 that in (2.3.2) we can choose $M=2^n$ and hence in (2.3.3) we have 
$$s=O_{\rho}\left(\ln n -\ln \epsilon \right).$$
We summarize the results of Sections 2.1-2.3 and Section 2.6.

\proclaim{(2.7) Theorem} Fix a $\rho > 1$. Let $\phi_i: {\Bbb R}^n \longrightarrow {\Bbb C}$ be functions such that $\phi_i$ is $L_i$-Lipschitz in the $\ell^1$ norm 
for some $L_i \geq 0$ and $i=1, \ldots, m$. For $j=1, \ldots, m$, let $I_j \subset \{1, \ldots, m\}$ be the set of indices $i$ such that $\phi_i$ depends on the coordinate 
$\xi_j$. Suppose that 
\roster
\item For some $r \geq 9$, each function $\phi_i$ depends on at most $r$ coordinates;
\item We have 
$$\sum_{i \in I_j} L_i \ \leq \ {1 \over 24 \rho \sqrt{r}} \quad \text{for} \quad j=1, \ldots, n;$$
\item We have 
$$\phi_i(0)=0 \quad \text{for} \quad i=1, \ldots, m.$$
\endroster
Then, for any 
 $0 < \epsilon < 1$, the expectation
$$\EE \exp\left\{ \sum_{i=1}^m \phi_i \right\}$$ 
with respect to the symmetric exponential measure with density (1.1.6)
can be approximated within relative error $\epsilon$  in polynomial time from the mixed moments 
$$\EE \left(\phi_{i_1} \cdots \phi_{i_l}\right) \quad \text{where} \quad 1 \leq i_1, \ldots, i_l \leq m, \quad l \leq s = O_{\rho} \left(\ln n - \ln \epsilon\right)$$
where the implied constant in the ``$O$" notation depends on $\rho$ only.
\endproclaim
{\hfill \hfill \hfill} \qed 

\head 3. Applications to counting integer points in polyhedra \endhead 

\subhead (3.1) Integer points in polyhedra \endsubhead 
Let ${\Bbb Z}^n_+ \subset {\Bbb R}^n$ be the set of points with non-negative integer coordinates, let $A=\left(a_{ij}\right)$ be a real $m \times n$ matrix and let $b=\left(b_1, \ldots, b_m \right)$ be a real $m$-vector. We define a polyhedron $P \subset {\Bbb R}^n$ by the system of linear equations and inequalities:
$$\aligned P=\Bigl\{ \left(\xi_1, \ldots, \xi_n\right): \quad &\sum_{j=1}^n a_{ij} \xi_j =b_i \quad \text{for} \quad i=1, \ldots, m \quad \text{and} \\
&\xi_j \ \geq \ 0 \quad \text{for} \quad j=1, \ldots, n \Bigr\}. \endaligned \tag3.1.1$$
Note that for any real $\sigma \ne 0$, the scaled matrix $\sigma A$ and vector $\sigma b$ define the same polyhedron $P$.

Suppose we want to compute the number $|P \cap {\Bbb Z}^n_+|$ of integer points in $P$. This is a well-known $\#$P-hard problem, and even deciding whether $P \cap {\Bbb Z}^n_+ \ne \emptyset$ is an NP-complete problem, so approximating $\left| P \cap {\Bbb Z}^n_+\right|$ is also computationally hard. Hence our goal is to find a reasonable computationally efficient relaxation of the problem. 

Let ${\bold q}=\left(q_1, \ldots, q_n\right)$ be a vector of real numbers $0 < q_j < 1$ for $j=1, \ldots, n$.
We consider the multivariate geometric distribution in ${\Bbb Z}^n_+$ defined by 
$${\Bbb P}(x) = \prod_{j=1}^n (1-q_j) q_j^{\xi_j} \quad \text{where} \quad x=\left(\xi_1, \ldots, \xi_n\right). \tag3.1.2$$
Note that 
$$\EE \xi_j = {q_j \over 1-q_j} \quad \text{for} \quad j=1, \ldots, n. \tag3.1.3$$

One reasonable choice of $q_j$ is the {\it maximum entropy distribution} \cite{BH10}, constructed as follows. Suppose that the polyhedron $P$ defined by (3.1.1) is bounded and has a non-empty relative interior, that is, contains a point $\left(\xi_1, \ldots, \xi_n\right)$ where 
$\xi_j > 0$ for all $j$. Then the strictly concave function 
$$g(x)=\sum_{j=1}^n \bigl(\left(\xi_j+1\right) \ln \left(\xi_j+1\right)-\xi_j \ln \xi_j \bigr) \quad \text{for} \quad x=\left(\xi_1, \ldots, \xi_n\right)$$
attains it maximum on $P$ at a unique point $z=\left(\zeta_1, \ldots, \zeta_n \right)$, which can be efficiently computed as a solution to a convex optimization problem. 
We necessarily have $\zeta_j > 0$ for all $j=1, \ldots, n$. Let us now define 
$$q_j={\zeta_j \over 1+\zeta_j} \quad \text{for} \quad j=1, \ldots, n.$$ 
It follows from (3.1.3) that 
$$\EE \left(\sum_{j=1}^n a_{ij} \xi_j \right) = b_i \quad \text{for} \quad i=1, \ldots, m,$$
so that the expectation of the random integer vector $x$ defined by (3.1.2) lies in $P$. Moreover,  see Theorem 4 in \cite{BH10}, we have $${\Bbb P}(x)=e^{-g(z)} \quad \text{for all} \quad x \in P \cap {\Bbb Z}^n_+,$$
so that the probability mass function is constant on the set of integer points in $P$
and hence 
$$|P \cap {\Bbb Z}^n_+| = e^{g(z)} {\Bbb P}\left(P \cap {\Bbb Z}^n_+\right). \tag3.1.4$$
Thus counting integer points in $P$ reduces to computing the probability in the right hand side of (3.1.4).

\subhead (3.2) Quadratic penalty function \endsubhead We want to replace the probability in the right hand side of (3.1.4) that is hard to compute by
an easier computable statistics.

For a matrix $A$ and a vector $b$ in (3.1.1), we define the {\it penalty function} $F_{A, b}: {\Bbb Z}^n_+ \longrightarrow {\Bbb R}_+$ by 
$$F_{A, b}(x)=\sum_{i=1}^m \left(-b_i + \sum_{j=1}^n a_{ij} \xi_j\right)^2 \quad \text{where} \quad x =\left(\xi_1, \ldots, \xi_n\right).$$
Hence $F_{A, b}(x)=0$ if $x$ satisfies the equations of (3.1.1) and $F_{A, b}(x) >0$ if $x$ does not. We then choose a vector ${\bold q}=\left(q_1, \ldots, q_n\right)$ where $0 < q_j < 1$ for $j=1, \ldots, n$ and consider the expectation of $\exp\left\{- {1 \over 2} F_{A, b} \right\}$ with respect to 
the multivariate geometric distribution (3.1.2):
$$\aligned &\EE \exp\left\{- {1 \over 2} F_{A, b} \right\} =\Phi(A, b; {\bold q}) \prod_{j=1}^n (1-q_j), \quad \text{where} \\
&\Phi\bigl(A, b; {\bold q}\bigr)=\sum\Sb x\in {\Bbb Z}^n_+ \\ x=\left(\xi_1, \ldots, \xi_n\right)\endSb \exp\left\{-{1 \over 2}\sum_{i=1}^m \left(-b_i+ \sum_{j=1}^n a_{ij} \xi_j\right)^2\right\} \prod_{j=1}^n q_j^{\xi_j}. \endaligned\tag3.2.1$$
Clearly, 
$$\Phi\bigl(A, b; {\bold q}\bigr) \prod_{j=1}^n (1-q_j) \ \geq \ {\Bbb P}\left(P \cap {\Bbb Z}^n\right),$$
so (3.2.1) provides an upper bound for (3.1.4), but unlike (3.1.4), it is amenable to an efficient approximation. We also note 
that 
$$\lim_{\sigma \longrightarrow +\infty} \Phi \bigl(\sigma A, \sigma b; {\bold q}\bigr) \prod_{j=1}^n (1-q_j)={\Bbb P}\left(P \cap {\Bbb Z}^n \right),$$
which ought to impose some limits on the computability of $\Phi(\sigma A, \sigma b; {\bold q})$ for large scaling factors $\sigma$, given the hardness of counting integer points.

Next, we fix $A$ and $b$ and extend $\Phi(A, b; {\bold q})$ defined by (3.2.1) to {\it complex} vectors ${\bold q}$ as a function ${\bold q} \longmapsto \Phi(A, b; {\bold q})$ of $q_j \in {\Bbb C}$ satisfying 
$$|q_j| < 1 \quad \text{for} \quad j=1, \ldots, n,$$
where we agree that $q_j^0=1$.
Clearly, $\Phi(A, b; {\bold q})$ is well-defined for such ${\bold q}$. It can be viewed as a partition function in the Potts model of a particular kind, cf. \cite{FV18}.

We have the following corollary of Theorem 1.2.
\proclaim{(3.3) Theorem} Let $A=\left(a_{ij}\right)$ be an $m \times n$ real matrix, let $b=\left(b_1, \ldots, b_m\right)$ be a real $m$-vector and let
${\bold q}=\left(q_1, \ldots, q_n\right)$ be a complex $n$-vector, such that $|q_j| < 1$ for $j=1, \ldots, n$.
Let $I_j \subset \{1, \ldots, m\}$, 
$$I_j = \bigl\{i: \ a_{ij} \ne 0 \bigr\} \quad \text{for} \quad j=1, \ldots, n$$ be the set of indices of rows with non-zero entries in the $j$-th column. Assume that 
\roster 
\item For some $r \geq 1$, we have 
$$|j: \ i \in I_j | \ \leq \ r \quad \text{for} \quad i=1, \ldots, m,$$
that is, each row of $A$ contains at most $r$ non-zero entries;
\item We have 
$$b_i^2 + \sum_{j =1}^n  {|q_j|^2 a_{ij}^2 \over (1-|q_j|)^2} \ \leq \ {1 \over 64 (r+1)} \quad \text{for all} \quad i=1, \ldots, m;$$
\item We have 
$$\sum_{i \in I_k} b_i^2 + \sum_{j:\ I_j \cap I_k \ne \emptyset} {|q_j|^2 \over (1-|q_j|)^2}  \sum_{i \in I_j} a_{ij}^2 \ \leq \ {1 \over 64 (r+1)} \quad \text{for all} \quad k=1, \ldots, n.$$
\endroster
Then for 
$$\Phi(A, b; {\bold q})=\sum\Sb x \in {\Bbb Z}^n_+ \\ x=\left(\xi_1, \ldots, \xi_n\right) \endSb \exp\left\{-{1 \over 2}\sum_{i=1}^m \left(-b_i+ \sum_{j=1}^n a_{ij} \xi_j\right)^2\right\} \prod_{j=1}^n  q_j^{\xi_j},$$
we have 
$$0 \ < \  \left| \Phi(A, b; {\bold q})\right| \ \leq \ \prod_{j=1}^n {1 \over 1-|q_j|}.$$
\endproclaim

We note that given ${\bold q}$, we can always scale $A \longmapsto \sigma A$ and $b \longmapsto \sigma b$ for $\sigma>0$, so that the polyhedron $P$ of (3.1.1) remains the same, while the scaled matrix $\sigma A$ and scaled vector $\sigma b$ satisfy the conditions of Theorem 3.3. 

Before we prove Theorem 3.3, we discuss how it leads to an efficient algorithm to approximate $\Phi(A, b; {\bold q})$. 

\subhead (3.4) Computing $\Phi(A, b, {\bold q})$ \endsubhead Let us fix $\rho >1$ and suppose that a matrix $A=\left(a_{ij}\right)$, a vector $b=\left(b_1, \ldots, b_m\right)$ and a vector $\rho {\bold q}=\left( \rho q_1, \ldots, \rho q_n\right)$ satisfy the conditions of Theorem 3.3. In particular, $|q_j|< \rho^{-1}$ for $j=1, \ldots, n$.
 Our goal is to approximate 
$\Phi(A, b; {\bold q})$.  

Let ${\Bbb D}_{\rho} \subset {\Bbb C}$ be the closed disc of radius $\rho$ in the complex plane, centered at 0. For given $A, b$ and ${\bold q}$, we define a function 
$g=g_{A, b, {\bold q}}: {\Bbb D}_{\rho} \longrightarrow {\Bbb C}$ by 
$$g(z)= \exp\left\{ {1 \over 2} \sum_{i=1}^m b_i^2 \right\} \Phi(A, b; z {\bold q}).$$
 Applying Theorem 3.3, we conclude that
$$g(0)=1 \quad \text{and} \quad 0\ < \  |g(z)|  \ \leq \ M \quad \text{for all} \quad z \in {\Bbb D}_{\rho},$$
where we can choose 
$$M= \exp\left\{ {1 \over 2} \sum_{i=1}^m b_i^2 \right\} \prod_{j=1}^n {1 \over 1-|q_j|} \ < \ e^{m/128} \left({\rho \over \rho-1}\right)^n.$$
It follows that one can approximate the value of 
$$\Phi(A, b; {\bold q})=\exp\left\{ -{1 \over 2} \sum_{i=1}^m b_i^2 \right\} g(1)$$ within relative error $0 < \epsilon < 1$ in polynomial time from the derivatives
$g^{(l)}(0)$ for $l=1, \ldots, s$, where 
$$s=O_{\rho} \left( \ln (n+m)  - \ln \epsilon\right), \tag3.4.1$$
cf. Section 2 and Theorem 2.5 in particular.
We have 
$$\split g^{(l)}(0)=&l! \exp\left\{{1 \over 2} \sum_{i=1}^m b_i^2 \right\} \\
&\qquad \times \sum\Sb (\xi_1, \ldots, \xi_n) \in {\Bbb Z}^n_+: \\ \xi_1 + \ldots + \xi_n = l \endSb \exp\left\{ -{1 \over 2} \sum_{i=1}^m \left(-b_i + \sum_{j=1}^n a_{ij} \xi_j \right)^2\right\} \prod_{j=1}^n q_j^{\xi_j}, \endsplit$$
and hence computing $g^{(l)}(0)$ reduces to the enumeration of ${n+l-1 \choose l}$ non-negative integer solutions to the equation $\xi_1 + \ldots + \xi_n=l$.

In view of (3.4.1), we obtain a quasi-polynomial deterministic algorithm to approximate $\Phi(A, b; {\bold q})$.
\bigskip
To get a feeling of the statistics computed by $\Phi(A, b; {\bold q})$, let us fix integers $r, c \geq 1$ and consider a family of instances, where each matrix $A=\left(a_{ij}\right)$ contains at most $r$ non-zero entries in each row and at most $c$ non-zero entries in each column. We allow the dimensions $m$ and $n$ of the matrix $A$ and vector $b$ to grow, but the sparsity parameters $r$ and $c$ remain fixed.
We suppose further that the entries $a_{ij}$ and $b_i$ are integer and stay uniformly bounded, $a_{ij}, b_i=O(1)$, and that the probabilities ${\bold q}=\left(q_1, \ldots, q_n\right)$ remain separated from 1, which by (3.1.3) enforces $\EE \xi_j=O(1)$, so that the expectations of the coordinates stay uniformly bounded.  Then there is a scaling $A \longmapsto \sigma A$, $b \longmapsto \sigma b$ that makes $A$ and $b$ satisfy the conditions of Theorem 3.3 with some fixed slack $\rho >1$ and hence makes $\Phi(A, b; {\bold q})$ efficiently computable, and such that the penalty $\left( b_i - \sum_{j=1}^n a_{ij}\xi_j\right)^2$ for violating an equation, when non-zero, is at least some positive constant $\Omega_{r, c}(1)$, depending on the sparsity parameters $r$ and $c$ alone. Hence the contribution of an $x\in {\Bbb Z}^n_+$ to $\Phi(A, b; {\bold q})$ is exponentially small in the number of violated equations in (3.1.1).
\bigskip
The proof of Theorem 3.3 is based on Theorem 1.2 and a Fourier transform trick, also known in statistical physics as the {\it Hubbard - Stratonovich transformation}, which allows one to toggle between Gaussian and discrete partition functions,  see Section 8.7.5 of \cite{FV18}.

\proclaim{(3.5) Lemma} Given an $m \times n$ real matrix $A=\left(a_{ij}\right)$, a real $m$-vector $b=\left(b_1, \ldots, b_m\right)$, and a complex $n$-vector
${\bold q}=\left(q_1, \ldots, q_n\right)$, where $|q_j| < 1$ for $j=1, \ldots, n$, we define $f_j: {\Bbb R}^m \longrightarrow {\Bbb C}$, $j=1, \ldots, n$,  by 
$$f_j(t)=\left(1-q_j \exp\left\{ \ii \sum_{i=1}^m a_{ij} \tau_i \right\}\right)^{-1} \quad \text{where} \quad t=\left(\tau_1, \ldots, \tau_m\right).$$
Then 
$$\Phi(A, b; {\bold q})={1 \over (2\pi)^{m/2}} \int_{{\Bbb R}^m} \exp\left\{-\ii \sum_{i=1}^m b_i \tau_i  -{1 \over 2} \sum_{i=1}^m \tau_i^2\right\} \prod_{j=1}^n f_j(t) \ dt.$$
\endproclaim 
\demo{Proof}
We use the well-known formula
$${1 \over \sqrt{2 \pi}} \int_{-\infty}^{+\infty} e^{\ii \alpha \tau} e^{-\tau^2/2} \ d \tau = \exp\left\{ - {\alpha^2\over 2}\right\}.$$

Let $t=\left( \tau_1, \ldots, \tau_m\right)$ be the standard Gaussian random $m$-vector and let $\EE$ denote the expectation with respect to the standard Gaussian probability measure in ${\Bbb R}^m$. Then
$$\split \Phi\bigl(A, b; {\bold q}\bigr)=&\sum\Sb x\in {\Bbb Z}^n_+ \\ x=\left(\xi_1, \ldots, \xi_n\right)\endSb \exp\left\{-{1 \over 2} \sum_{i=1}^m \left(-b_i+ \sum_{j=1}^n a_{ij} \xi_j\right)^2\right\} \prod_{j=1}^n q_j^{\xi_j}.\\
=&\sum\Sb x\in {\Bbb Z}^n_+ \\ x=\left(\xi_1, \ldots, \xi_n\right)\endSb \left(\EE \exp\left\{ \ii \sum_{i=1}^m \tau_i \left(-b_i + \sum_{j=1}^n a_{ij} \xi_j \right)\right\}\right)\prod_{j=1}^n q_j^{\xi_j} \\ 
=&\EE \Biggl( \exp\left\{ -\ii \sum_{i=1}^m b_i \tau_i\right\} \\ &\qquad \times \sum\Sb x\in {\Bbb Z}^n_+ \\ x=\left(\xi_1, \ldots, \xi_n\right)\endSb \exp\left\{ \ii \sum_{j=1}^m \xi_j \left(\sum_{i=1}^m a_{ij} \tau_i\right)\right\}  \prod_{j=1}^n q_j^{\xi_j}  \Biggr)\\
=&\EE \left( \exp\left\{ -\ii \sum_{j=1}^m b_i \tau_i \right\} \prod_{j=1}^n f_j(t) \right),\endsplit$$
as required.
{\hfill \hfill \hfill} \qed 
\enddemo

\subhead (3.6) Proof of Theorem 3.3 \endsubhead We define functions $\phi_k: {\Bbb R}^m \longrightarrow {\Bbb C}$, $k=1, \ldots, m+n$,  as follows.
For $t \in {\Bbb R}^m$, $t=\left(\tau_1, \ldots, \tau_m \right)$, we define
$$\phi_i(t)=-\ii b_i \tau_i \quad \text{for} \quad i=1, \ldots, m$$
and
$$\phi_{m+j}(t)=-\ln \left(1 -q_j \exp\left\{ \ii \sum_{i=1}^m a_{ij} \tau_i \right\} \right) \quad \text{for} \quad j=1, \ldots, n. $$
Since $|q_j| < 1$, the functions $\phi_{m+j}$ are well-defined by the choice of the branch of the logarithm for $\phi_{m+j}(0)=-\ln (1-q_j)$.  By Lemma 3.5, 
$$\Phi(A, b; {\bold q})=\EE \exp\left\{  \sum_{i=1}^{m+n} \phi_i \right\}, \tag3.6.1$$
where the expectation is taken with respect to the standard Gaussian probability measure in ${\Bbb R}^m$. 

To apply Theorem 1.2, we bound the Lipschitz constants of $\phi_i$ in the $\ell^2$ norm of ${\Bbb R}^m$. Clearly, for $i=1, \ldots, m$, the function $\phi_i$ is $|b_i|$-Lipschitz. Computing the gradient of $\phi_{m+j}$, we get 
$${\partial \phi_{m+j} \over \partial \tau_i} = \ii  a_{ij} q_j {\exp\left\{\ii \sum_{i=1}^m a_{ij} \tau_i \right\} \over 1-q_j \exp\left\{\ii \sum_{i=1}^m a_{ij} \tau_i \right\}}\quad \text{for} \quad 
j=1, \ldots, n$$
and hence the Lipschitz constant of $\phi_{m+j}$ does not exceed
$$\| \nabla \phi_{m+j} \|_2 \ \leq \ {|q_j| \over 1 - |q_j|} \left( \sum_{i=1}^m a_{ij}^2\right)^{1/2}.$$
Next, we observe that for each $i$, at most $r+1$ functions $\phi_k$ depend on the coordinate $\tau_i$. 
Each function $\phi_i(t)$ shares a coordinate with functions $\phi_{m+j}(t)$ if and only if $i \in I_j$. Similarly, each function $\phi_{m+k}(t)$ shares a coordinate 
with functions $\phi_i(t)$ with $i \in I_k$ and with functions $\phi_{m+j}(t)$ such that $I_k \cap I_j \ne \emptyset$.

 Applying Theorem 1.2 to (3.6.1), we conclude that 
$\Phi(A, b; {\bold q}) \ne 0$.
Clearly,
$$|\Phi(A, b; {\bold q})| \ \leq \ \sum\Sb x \in {\Bbb Z}^n_+ \\ x=\left(\xi_1, \ldots, \xi_n \right) \endSb \prod_{j=1}^n |q_j|^{\xi_j} = \prod_{j=1}^n {1 \over 1-|q_j|}.$$
{\hfill \hfill \hfil} \qed 

\head 4. Applications to volume computation \endhead

Randomized Markov Chain Monte Carlo algorithms have been spectacularly successful in efficiently approximating volumes of convex bodies, see \cite{LV07}. Deterministic algorithms were noticeably less so. Approximating the volume of a general oracle-defined convex body is provably difficult for deterministic algorithms 
\cite{BF87}. This, however, leaves a theoretical possibility of efficient deterministic algorithms for structured sets, such as polyhedra, and indeed there were some recent successful attempts for some special combinatorially defined polytopes 
\cite{BR26}, \cite{GN25}.

In this section, we apply Theorem 1.3 to approximate volumes of compact sets of a particular structure. Loosely, they can be described as perturbations of cross-polytopes (higher-dimensional octahedra). Some of those sets are non-convex.

The application is based on a simple formula.

\proclaim{(4.1) Lemma} Let $\Psi: {\Bbb R}^n \longrightarrow {\Bbb R}_+$ be a continuous function such that 
$$\Psi(\alpha x) = \alpha \Psi(x) \quad \text{for all} \quad x \in {\Bbb R}^n \quad \text{and all} \quad \alpha \geq 0$$
and 
$$\Psi(x) = 0 \Longrightarrow x=0.$$
Then for the set $K_{\Psi} \subset {\Bbb R}^n$, 
$$K_{\Psi}=\Bigl\{x: \ \Psi(x) \leq 1 \Bigr\},$$
we have 
$$\vl K_{\Psi}= {1 \over n!} \int_{{\Bbb R}^n} e^{-\Psi(x)} \ dx.$$
\endproclaim
\demo{Proof} Clearly, $K_{\Psi}$ is a compact set.
For $t >0$, we have
$$t K_{\Psi} =\Bigl\{ x: \ \Psi(x) \leq t \Bigr\}.$$
Let
$$F(t)=\vl\left(t K_{\Psi}\right) = t^n \vl (K) \quad \text{for} \quad t >0.$$
Then 
$$\int_{{\Bbb R}^n} e^{-\Psi(x)} \ dx = \int_0^{+\infty} e^{-t} \ d F(t)= \vl (K) \int_0^{+\infty} n t^{n-1} e^{-t} \ d t = n! \vl(K).$$
{\hfill \hfill \hfill} \qed 
\enddemo

\subhead (4.2) Examples \endsubhead
Let us fix some $\rho>1$.
Suppose that 
$$\Psi(x)=\sum_{j=1}^n |\xi_j| +\sum_{i=1}^m \phi_i(x) \quad \text{for} \quad x=\left(\xi_1, \ldots, \xi_n\right),$$
where $\phi_i: {\Bbb R}^n \longrightarrow {\Bbb R}_+$ satisfy the conditions of Theorem 2.7 with the slack $\rho$ and, in addition, $\phi_i$ are positive homogeneous of degree $1$:
$$\phi_i(\alpha x)= \alpha \phi_i(x) \quad \text{for} \quad \alpha \geq 0 \quad \text{and}\quad i=1, \ldots, m.$$
By Lemma 4.1, we have 
$$\vl K_{\Psi}= {1 \over n!} \int_{{\Bbb R}^n} e^{-\Psi(x)} \ dx, \tag4.2.1$$
and Theorem 2.7 implies that to approximate (4.2.1) within relative error $0 < \epsilon < 1$, one needs to compute
$m^{O_{\rho}(\ln n-\ln \epsilon)}$ integrals 
$$\split &\int_{{\Bbb R}^n} \phi_{i_1}(x) \cdots \phi_{i_l}(x) \exp\left\{ - \sum_{j=1}^n |\xi_j| \right\} \ dx \\ &\text{for} \quad l \leq s=O_{\rho}\left(\ln n -\ln \epsilon\right).\endsplit
\tag4.2.2$$

One natural example of such functions $\phi_i$ is provided by {\it support functions} of convex bodies, and, in particular, polytopes. Namely, let $B_i \subset {\Bbb R}^n$ be a convex body containing the origin in its relative interior, and suppose that 
$$\phi_i(x)= \max_{y \in B_i} \langle x, y \rangle,$$
where $\langle \cdot, \cdot \rangle$ is the standard scalar product in ${\Bbb R}^n$. Clearly, $\phi_i$ is positive homogeneous of degree $1$. Since the function 
$$x \longrightarrow \sum_{j=1}^n |\xi_j| \quad \text{for} \quad x=\left(\xi_1, \ldots, \xi_n\right)$$
is the support function of the cube $C_n=[-1, 1]^n$, then the set $K_{\Psi}$ is the polar of the Minkowski sum,
$$K_{\Psi} = \left( C_n + \sum_{i=1}^m B_i \right)^{\circ}. \tag4.2.3$$
If $B_i$ lies in a coordinate subspace of dimension $r_i$ then $\phi_i$ depends on at most $r_i$ coordinates. Moreover, if $B_i$ is a polytope defined as the convex hull of $v_i$ vertices, then 
$\phi_i$ are a piece-wise linear functions, and the integrals (4.2.2) can be computed in $(rv)^{O(r)}$ time, where $r=r_{i_1}+ \ldots + r_{i_l}$ and $v=v_{i_1}+ \ldots + v_{i_l}$, 
see for example, \cite{B+11}.

We remark that if $B_i$ is an interval, then the Minkowski sum $C_n +  \sum_{i=1}^m B_i$ is a Minkowski sum of intervals, that is a {\it zonotope}. 
Assuming that $B_i=-B_i$, we can write $C_n + \sum_{i=1}^m B_i$ as a linear image (projection) of the cube $C_{m+n}$ in ${\Bbb R}^{m+n}$. Then the polar $K_{\Psi}$ defined by (4.2.3) is interpreted as a section of a linear image of the {\it cross-polytope}, that is, the unit ball in the $\ell^1$ norm in ${\Bbb R}^{m+n}$.
\bigskip
Sets $K_{\Psi}$ for which $\vl K_{\Psi}$ can be efficiently approximated via Theorem 2.7 include some non-convex sets. Suppose, for example, that 
$\Psi: {\Bbb R}^2 \longrightarrow {\Bbb R}_+$ is defined by 
$$\Psi(x) = |\xi_1| + |\xi_2| - \epsilon |\xi_1 - \xi_2| \quad \text{where} \quad x=\left(\xi_1, \xi_2\right)$$
and $0 < \epsilon < 1$. Then the points $(1+\epsilon, 0)$ and $(0, 1+\epsilon)$ lie in $K_{\Psi}$ but their average $\left({1+\epsilon \over 2}, {1+\epsilon \over 2}\right)$ does not.

\head 5. Proof of Theorem 1.2 \endhead 

To simplify notation, in this section we denote the $\ell^2$ norm in ${\Bbb R}^n$ just by $\| \cdot \|$. We denote the standard Gaussian measure with density (1.1.5) by $\mu$ and denote the expectation with respect to that measure by $\EE$. We say that a function $f: {\Bbb R}^n \longrightarrow {\Bbb C}$ is $L$-Lipschitz, if it is $L$-Lipschitz in the $\ell^2$ norm $\| \cdot\|$.

Our proof is based on the following well-known result regarding real-valued Lipschitz functions.
\proclaim{(5.1) Theorem} Let $f: {\Bbb R}^n \longrightarrow {\Bbb R}$ be an $L$-Lipschitz function such that $\EE f=0$.
Then
\roster 
\item We have
$$\EE e^f \ \leq \ \exp\left\{ {L^2 \over 2} \right\}; $$
\item For $a \geq 0$, we have
$$\mu\Bigl\{ x \in {\Bbb R}^n: \quad f(x) \geq a \Bigr\} \ \leq \ \exp\left\{ -{a^2 \over 2 L^2}\right\}.$$
\endroster
\endproclaim
\demo{Proof} See, for example, Section 5.1 of \cite{Le01}.
{\hfill \hfill \hfill} \qed
\enddemo

Next, we turn to complex-valued Lipschitz functions. 
\proclaim{(5.2) Lemma} Let $f: {\Bbb R}^n \longrightarrow {\Bbb C}$ be an $L$-Lipschitz function for $L={1 \over 4}$.
Then 
$$\left| \EE e^f \right| \ \geq \ {1 \over 2} \EE \left| e^f \right|.$$
\endproclaim
\demo{Proof} Without loss of generality, we assume that $\EE f=0$. Let $f=g+\ii h$, where $g, h: {\Bbb R}^n \longrightarrow {\Bbb R}$ are 
real-valued $L$-Lipschitz functions such that $\EE g=\EE h =0$. From the Jensen inequality, we obtain
$$\EE e^g \ \geq \ 1. \tag5.2.1$$
We have 
$$e^{f(x)} = e^{g(x)} \left( \cos h(x) + \ii \sin h(x)\right) \quad \text{for all} \quad x \in {\Bbb R}^n.$$

Let
$$X=\Bigl\{ x \in {\Bbb R}^n: \quad |h(x)| \ \leq \ 1 \Bigr\} \quad \text{and} \quad \overline{X}={\Bbb R}^n \setminus X.$$ From Part (2) of Lemma 5.1, we have
$$\mu(\overline{X})\ \leq \ 2\exp\left\{ - {1 \over 2L^2}\right\}=2e^{-8}. \tag5.2.2$$
For all $x \in X$, we have $|h(x)| \leq 1$ and hence 
$$\aligned \left| \int_{X} e^f \ d \mu \right| \ \geq \ &\Re\thinspace \left( \int_{X} e^f \ d \mu \right)= \int_{X} \Re\thinspace e^f \ d \mu \ \geq \ \int_{X} (\cos 1) e^g \ d \mu \\ =& \ (\cos 1) \int_{X} e^g \ d \mu. \endaligned \tag5.2.3$$
Let $[\overline{X}]$ be the indicator of $\overline{X}$, that is, the function $[\overline{X}]: {\Bbb R}^n \longrightarrow {\Bbb R}$ such that 
$$[\overline{X}](x)=\cases 1 &\text{if\ } x \notin X \\ 0 &\text{if\ } x \in X. \endcases$$
We use the H\"older inequality with 
$$p={1 \over L^2} = 16 \quad \text{and} \quad q={1 \over 1-L^2} = {16 \over 15},$$
and get 
$$\int_{\overline{X}} e^g \ d \mu = \EE\left( [\overline{X}] e^g \right) \ \leq \ \left( \mu(\overline{X})\right)^{1/q} \left(\EE e^{pg}\right)^{1/p}.$$
Since $pg$ is $pL$-Lipschitz, from Part (1) of Lemma 5.1, we have 
$$\EE e^{pg} \ \leq \ \exp\left\{ {p^2 L^2 \over 2} \right\}$$
and hence from (5.2.2) we obtain
$$\aligned &\int_{\overline{X}} e^g \ d \mu \ \leq \ \left(2 \exp\left\{ - {1 \over 2L^2}\right\} \right)^{1/q} \exp\left\{ {p L^2 \over 2} \right\} \\
&\qquad < \ 2 \exp\left\{-{1-L^2 \over 2 L^2} + {1 \over 2} \right\} = 2 \exp\left\{ 1 - {1 \over 2L^2} \right\} =2 e^{-7}.\endaligned  \tag5.2.4$$

Let
$$a=\int_X e^g \ d \mu \quad \text{and} \quad b = \int_{\overline{X}} e^g\ d \mu.$$
Clearly, 
$$\EE \left| e^f \right| =\EE e^g = a+b$$
and from (5.2.3), we obtain 
$$\left| \EE e^f \right| \ \geq \ a (\cos 1) - b.$$ Therefore,
$${\left| \EE e^f \right| \over \EE \left| e^f \right| } \ \geq \ {a (\cos 1) - b \over a+b} = {(\cos 1) - (b/a) \over 1+ (b/a)}.$$
From (5.2.1), we have $a+b \geq 1$ and hence $a \geq 1- b$. Therefore, 
$${b \over a} \ \leq \ {b \over 1-b}$$
and 
$${\left| \EE e^f \right| \over \EE \left| e^f \right| } \ \geq \ {(\cos 1) - b/(1-b) \over 1+ b/(1-b)} = (\cos 1) (1-b) -b.$$
Since by (5.2.4), we have  $b <  2 e^{-7}$, we finally obtain 
$${\left| \EE e^f \right| \over \EE \left| e^f \right| } \ \geq \ (\cos 1) \left(1 -2 e^{-7}\right) - 2 e^{-7} \approx 0.5374931582 \ > \ 0.5,$$
as desired.
{\hfill \hfill \hfill} \qed 
\enddemo

Now we are ready to prove Theorem 1.2.

We prove by induction on the number $m$ of functions the following stronger statement.
\bigskip
 {\bf (5.3) Claim:}  Let $\phi_1, \ldots, \phi_m: {\Bbb R}^n \longrightarrow {\Bbb C}$ be functions and let $L_1, \ldots, L_m \geq 0$  be their respective Lipschitz constants, as in Theorem 1.2. Let $\widehat{\phi}_m: {\Bbb R}^n \longrightarrow {\Bbb C}$ be yet another function, which is $L_m$-Lipschitz 
and depends on a subset or all of the coordinates that $\phi_m$ depends on. Suppose further that 
$$\left| \widehat{\phi}_m(x) -\phi_m(x) \right| \ \leq \ \tau \quad \text{for all} \quad x \in {\Bbb R}^n \tag5.3.1$$
and some $\tau >0$. 
Then we have 
$$\EE \exp\left\{ \phi_m + \sum_{i=1}^{m-1}  \phi_i \right\} \ne 0, \quad \EE \exp\left\{  \widehat{\phi}_m + \sum_{i=1}^{m-1}\phi_i \right\} \ne 0 \tag5.3.2$$
and the ratio of the two numbers (5.3.2) can be written as $e^{\alpha}$ for some $\alpha \in {\Bbb C}$ such that 
$$|\alpha| \ \leq \ 2 \tau. \tag5.3.3$$
\bigskip
{\bf (5.4) Base $m=1$.} To simplify the notation somewhat, we drop the index 1 and denote the functions in question just by $\phi$ and $\widehat{\phi}$ and number $L_1$ just by $L$.
For $0 \leq s \leq 1$, let
$$\tilde{\phi}_s =(1-s) \phi + s \widehat{\phi}.$$
Hence $\tilde{\phi}_s$ is $L$-Lipschitz, $\tilde{\phi}_0=\phi$ and $\tilde{\phi}_1=\widehat{\phi}$. Since $L \leq 1/8 < 1/4$, by Lemma 5.2 we have  $\EE e^{\tilde{\phi}_s} \ne 0$ and we can choose a continuous branch of the function $s \longrightarrow \ln \EE e^{\tilde{\phi}_s}$ in some neighborhood of the interval $[0, 1] \subset {\Bbb C}$.
Now,
$$\ln \EE e^{\widehat{\phi}} -\ln \EE e^{ \phi} = \int_0^1\left( {d \over ds} \ln \EE e^{ \tilde{\phi}_s} \right) \ ds =
\int_0^1  {\EE ( \widehat{\phi}-\phi) e^{\tilde{\phi}_s} \over \EE e^{ \tilde{\phi}_s}} \ ds. \tag5.4.1 $$
We have 
$$\left| \EE  ( \widehat{\phi}-\phi) e^{\lambda \tilde{\phi}_s} \right| \ \leq \ \EE | \widehat{\phi}-\phi| \left| e^{\tilde{\phi}_s} \right| \ \leq \ 
\tau \EE \left| e^{ \tilde{\phi}_s} \right|.$$
Since $L \leq 1/8 < 1/4$, by Lemma 5.2, we have 
$$\left| \EE e^{\tilde{\phi}_s} \right| \ \geq \ {1 \over 2} \EE \left| e^{ \tilde{\phi}_s}\right|.$$
Therefore, from (5.4.1),
$$\left| \ln \EE e^{ \widehat{\phi}} -\ln \EE e^{ \phi} \right| \ \leq \ 2 \tau,$$
and Claim 5.3 follows.
\bigskip
{\bf (5.5) Induction step $m-1 \Longrightarrow m$ for $m \geq 2$.} Let $J \subset \{1, \ldots, n\}$ be the set of indices $j$ such that $\phi_m$ depends on $\xi_j$ and let $\overline{J}=\{1, \ldots, n\} \setminus J$ be its complement. 

If $J=\emptyset$ then $\phi_m$ and $\widehat{\phi}_m$ are constants and the ratio of the two expectations 
(5.3.2) can be written as $e^{\alpha}$ for $\alpha=\pm  \left(\widehat{\phi}_m - \phi_m \right)$
where the sign is chosen depending on the order in which the ratio is taken, and hence $|\alpha| \leq  \tau$ from (5.3.1).

Suppose that $\overline{J}=\emptyset$, so that every function $\phi_i$ shares a coordinate with $\phi_m$ and hence $I_m=\{1, \ldots, m\}$.
Similarly to the Section 5.4, for $0 \leq s \leq 1$, we introduce a function $f_s: {\Bbb R}^n \longrightarrow {\Bbb C}$ 
 by 
$$f_s= (1-s) \phi_m + s\widehat{\phi}_m  + \sum_{i=1}^{m-1}  \phi_i.$$ 
For $i=1, \ldots, m$, let $\left(\xi_j: \ j \in J_i\right)$ be the coordinates that $\phi_i$ depends on, and for $x \in {\Bbb R}^n$, let $x_i \in {\Bbb R}^{J_i}$ be the vector of those coordinates. Since the function $(1-s) \phi_m + s \widehat{\phi}_m$ is $L_m$-Lipschitz and the functions $\phi_i$ for $i=1, \ldots, m-1$ are $L_i$-Lipschitz, via the Cauchy - Schwarz inequality, for all $x, y \in {\Bbb R}^n$, we obtain
$$\split & |f_s(x)-f_s(y)| \ \leq \ L_m \left\| x_m - y_m \right\|+ \sum_{i=1}^{m-1} L_i \left\| x_i -y_i  \right\|  \\ &\qquad \leq \ \left(\sum_{i=1}^m L_i^2 \right)^{1/2} \left( \sum_{i=1}^m \left\| x_i -y_i \right\|^2 \right)^{1/2} \ \leq \ \left({1 \over 64 c} \right)^{1/2}  c^{1/2} \| x-y\| \\ &\qquad = {1 \over 8} \|x-y\|.\endsplit$$
(Here to bound $\sum_{i=1}^m \|x_i-y_i\|^2$ we used that for every $j$ there are at most $c$ sets $J_i$ with $j \in J_i$.)

Hence $f_s$ are ${1 \over 8}$-Lipschitz and by Lemma 5.2 we have 
$$\left| \EE e^{f_s} \right| \ \geq \ {1 \over 2} \left| \EE e^{f_s} \right| \ > \ 0 \quad \text{for all} \quad 0 \leq s \leq 1.$$
Then arguing exactly as in Section 5.4, we write 
$$\split \left| \ln \EE e^{f_1} - \ln \EE e^{f_0}\right| = &\left| \int_0^1 \left({d \over ds} \ln \EE e^{f_s}\right)\ ds\right| =
\left|  \int_0^1 { \EE \left(\widehat{\phi}_m-\phi_m \right) e^{f_s} \over \EE e^{f_s}}  \ ds \right| 
\\ \leq \ & \tau \int_0^1 {\EE \left| e^{f_s}\right| \over \left| \EE e^{f_s} \right|} \ ds \ \leq \ 2 \tau, \endsplit$$
as required by (5.3.3).

Thus without loss of generality, we assume that $J \subset \{1, \ldots, n\}$ is a proper subset. 
 We represent ${\Bbb R}^n$ as the direct sum ${\Bbb R}^n = {\Bbb R}^J \oplus {\Bbb R}^{\overline{J}}$. Let $I=I_m \setminus \{m\}$, so that $I$ consists of the indices $i \ne m$ such that $\phi_i$ shares a coordinate with $\phi_m$. 
For a vector $x \in {\Bbb R}^J$ and a function $\phi_i$ with $i \in I$, we define a function $\phi_i(\cdot\ | x): {\Bbb R}^{\overline{J}}
\longrightarrow {\Bbb C}$ obtained by restricting the coordinates $\xi_j$ with $j \in J$ to those of $x$. Further, we define 
$\Psi: {\Bbb R}^J \longrightarrow {\Bbb C}$ by 
$$\Psi(x) = \EE_{\overline{J}} \exp\left\{ \sum_{i \in I} \phi_i(\cdot\ | x ) +\sum\Sb 1 \leq i \leq m-1 \\ i \notin I \endSb \phi_i \right\}, \tag5.5.1$$
where the expectation is taken with respect to the coordinates $\xi_j$ with $j \in \overline{J}$. 

We compare values $\Psi(x')$ and $\Psi(x'')$ for two vectors 
$x', x'' \in {\Bbb R}^J$. Switching from $x' =\left(\xi'_j:\ j \in J\right)$ to $x''=\left(\xi''_j:\ j \in J\right)$ in (5.5.1) affects the functions $\phi_i(\cdot\ | x)$ for $i \in I$. 
For $i \in I_m=I \cup \{m\}$, let $J_i \subset J$ be the set of indices $j \in J$ of the coordinates $\xi_j$ shared by $\phi_i$ and $\phi_m$, and for $x \in {\Bbb R}^J$, let $x_i$ be the vector of the coordinates $\xi_j \in {\Bbb R}^{J_i}$ for $j \in J_i$. In particular, $J_m=J$ and $x_m=x$.
We have 
$$\left| \phi_i (\cdot\ | x') - \phi_i (\cdot\ | x'') \right| \ \leq \ L_i \left\| x'_i-x''_i\right\| \quad \text{for} \quad i \in I_m.$$
Applying the induction hypothesis, Claim 5.3, $|I|$ times, we conclude that $\Psi(x') \ne 0$, $\Psi(x'') \ne 0$ and that we can write 
$${\Psi(x') \over \Psi(x'')} = e^{\alpha} \qquad \text{for some} \quad \alpha \in {\Bbb C} \quad \text{satisfying} \qquad  |\alpha| \ \leq \ 
2 \sum_{i \in I} L_i  \left\| x'_i - x''_i \right\|.$$
It follows now that we can choose a continuous branch 
$$\psi: {\Bbb R}^J \longrightarrow {\Bbb C}, \quad \psi(x)= \ln \Psi(x),$$
such that 
$$\left| \psi(x') - \psi(x'') \right| \ \leq \ 2  \sum_{i \in I} L_i \| x_i' -x_i'' \| \quad \text{for all} \quad x', x'' \in {\Bbb R}^J. \tag5.5.2$$

We have 
$$\aligned &\EE \exp\left\{ \phi_m+  \sum_{i=1}^{m-1} \phi_i \right\} = \EE_J \exp\left\{ \phi_m +\psi \right\} \quad \text{and, similarly,} \\
&\EE  \exp\left\{ \widehat{\phi}_m + \sum_{i=1}^{m-1} \phi_i \right\} = \EE_J  \exp\left\{ \widehat{\phi}_m + \psi \right\}, \endaligned \tag5.5.3 $$
where the expectation in the right hand side is taken with respect to the coordinates $\xi_j$ with $j \in J$.

As in Section 5.4, for $0 \leq s \leq 1$, we define 
$$\tilde{\phi}_s =(1-s) \phi_m + s \widehat{\phi}_m,$$
so that $\tilde{\phi}_0=\phi_m$, $\tilde{\phi}_1= \widehat{\phi}_m$, and $\tilde{\phi}_s$ is $L_m$-Lipschitz. Using (5.5.2) and the Cauchy - Schwarz inequality, we conclude
that 
$$\aligned &\left| \left( \tilde{\phi}_s(x') + \psi(x') \right) - \left( \tilde{\phi}_s(x'') + \psi(x'') \right) \right| \\ &\qquad \leq \ L_m \|x'-x''\| + 
2 \sum_{i \in I} L_i \|x'_i-x_i''\| \ \leq \ 2 \sum_{i \in I_m} L_i \|x_i' -x_i''\| \\ &\qquad \ \leq  \ 
2 \left( \sum_{i \in I_m} L_i^2 \right)^{1/2} \left( \sum_{i \in I_m} \|x_i'-x_i''\|^2 \right)^{1/2}.\endaligned \tag5.5.4$$
Now, for each coordinate $\xi_j$ with $j \in J$, there are at most $c$ functions $\phi_i$ with $i \in I_m$ that depend on $\xi_j$. Therefore, 
$$\sum_{i \in I_m} \left\|x_i'-x_i'' \right\|^2 \ \leq \ c \|x'-x''\|^2. \tag5.5.5$$
Combining (5.5.4) and (5.5.5), we obtain
$$\split &\left| \left(  \tilde{\phi}_s(x') + \psi(x') \right) - \left( \tilde{\phi}_s(x'') + \psi(x'') \right) \right| \ \leq \ 2 \sqrt{c}  \left(\sum_{i \in I_m} L_i^2
\right)^{1/2} \|x'-x''\| \\
&\qquad \leq \ 2 c^{1/2} \left({1 \over 64 c}\right)^{1/2}  \|x'-x''\| ={1 \over 4} \|x'-x''\|. \endsplit $$
Hence we conclude that for any $s$ the function $ \tilde{\phi}_s + \psi$ is ${1 \over 4}$-Lipschitz.

By Lemma 5.2, we have 
$$\left| \EE_J \exp\left\{ \tilde \phi_s + \psi \right\}\right| \ \geq \ {1 \over 2} \EE_J \left| \exp\left\{ \tilde \phi_s + \psi \right\}\right| \ > \ 0.
\tag5.5.6$$
We choose a continuous branch of the function 
$s \longmapsto \ln \EE_J \exp\left\{ \tilde{\phi}_s + \psi \right\}$ in some neighborhood of $[0, 1] \subset {\Bbb C}$ and argue as before. Using (5.5.6), we obtain
$$\split &\left| \ln \EE_J \exp\left\{  \widehat{\phi}_m + \psi \right\} - \ln \EE_J \exp\left\{\phi_m + \psi \right\} \right| \\&\qquad =\left| \int_0^1 \left( {d \over ds} 
\ln \EE_J \exp\left\{ \tilde{\phi}_s + \psi \right\} \right) \ ds \right| \\
&\qquad = \left| \int_0^1  {\EE_J (\widehat{\phi}_m -\phi_m) \exp\left\{\tilde{\phi}_s + \psi\right\}  \over \EE_J  \exp\left\{ \tilde{\phi}_s + \psi\right\} } \ ds \right| \\
&\qquad \ \leq \ \tau \int_0^1 {\EE_J  \left| \exp\left\{ \tilde{\phi}_s + \psi \right\} \right| \over \left| \EE_J \exp\left\{ \tilde{\phi}_s \right\} \right|} \ ds \\
&\qquad \ \leq \ 2 \tau.\endsplit  \tag5.5.7$$
The proof of Claim 3.1 now follows by (5.5.3) and (5.5.7).
{\hfill \hfill \hfill} \qed

\head 6. Proof of Theorem 1.3 \endhead 

To simplify notation, in this section we denote the $\ell^1$ norm in ${\Bbb R}^n$ just by $\| \cdot \|$. We denote the standard exponential measure with density (1.1.6) by $\mu$ and denote the expectation with respect to that measure by $\EE$. We say that a function $f: {\Bbb R}^n \longrightarrow {\Bbb C}$ is $L$-Lipschitz, if it is $L$-Lipschitz in the $\ell^1$ norm $\| \cdot \|$. 

The proof is very similar to the proof of Theorem 1.2 in Section 5. First, we summarize some results regarding the symmetric exponential measure and real-valued Lipschitz functions.

\proclaim{(6.1) Theorem} Suppose that $F: {\Bbb R}^n \longrightarrow {\Bbb R}$ is a differentiable function such that $\EE F=0$, 
$$\max_{j=1, \ldots, n} \left| {\partial F \over \partial \xi_j} \right| \ \leq \ 1, \quad x=\left(\xi_1, \ldots, \xi_n\right),$$
and 
$$\sum_{j=1}^n \left( {\partial F \over \partial \xi_j}\right)^2 \ \leq \ a^2 \quad \text{for some} \quad a > 0.$$
Then
\roster
\item For $-{1 \over 2} \ \leq \ \lambda \ \leq \ {1 \over 2}$ we have
$$\EE e^{\lambda F} \ \leq \ \exp\left\{ 4 a^2 \lambda^2 \right\};$$
\item For $r \geq 0$ we have
$$\mu\Bigl\{ x \in {\Bbb R}^n: \quad F(x) \ \geq \ r  \Bigr\} \ \leq \ \exp\left\{ -{1 \over 4} \min\left\{ r, {r^2 \over 4a^2}\right\}\right\}.$$
\endroster
\endproclaim
\demo{Proof} See Section 5.3, in particular p. 105 of \cite{Le01}.
{\hfill \hfill \hfill} \qed 
\enddemo

Next, we turn to complex-valued Lipschitz functions.

\proclaim{(6.2) Lemma} Suppose that $f: {\Bbb R}^n \longrightarrow {\Bbb C}$ is $L$-Lipschitz for 
$$L=\min\left\{{1 \over 12 \sqrt{n}}, \ {1 \over 36} \right\}.$$
Then 
$$\left| \EE e^f \right| \ \geq \ {1 \over 2}  \EE \left| e^f \right|.$$
\endproclaim
\demo{Proof} A standard argument allows us to assume that $f$ is differentiable. Indeed, for $\sigma >0$, let 
$\psi_{\sigma}: {\Bbb R}^n \longrightarrow {\Bbb R}_+$ be the Gaussian density with the covariance matrix  $\sigma I$, 
$$\psi_{\sigma}(x)={1 \over (2\pi \sigma^2)^{n/2}} \exp\left\{ -{1 \over 2 \sigma^2} \sum_{j=1}^n \xi_j^2 \right\} \quad \text{for} \quad x=\left(\xi_1, \ldots, \xi_n \right),$$
and let $f_{\sigma}: {\Bbb R}^n \longrightarrow {\Bbb C}$ be the convolution of $f$ and $\psi_{\sigma}$:
$$f_{\sigma}(x)=\int_{{\Bbb R}^n} f(x-y) \psi_{\sigma}(y) \ dy = \int_{{\Bbb R}^n} f(y) \psi_{\sigma}(x-y) \ d y.$$
From the first integral representation it follows that $f_{\sigma}$ is $L$-Lipschitz and from the second representation it follows that $f_{\sigma}$ is differentiable. In addition, $f_{\sigma} \longrightarrow f$ uniformly on compact sets, as $\sigma \longrightarrow 0+$.

Next, without loss of generality we assume that $\EE f=0$. 

Let $f=g+\ii h$, where $g, h: {\Bbb R}^n \longrightarrow {\Bbb R}$ are $L$-Lipschitz such that $\EE g=\EE h=0$. 
From the Jensen inequality, we obtain 
$$\EE e^g \ \geq \ 1. $$
In addition, 
$$\left| {\partial g \over \partial \xi_j} \right|, \quad \left| {\partial h \over \partial \xi_j} \right| \ \leq \ L \quad \text{for}  \quad j=1, \ldots, n.$$
Let 
$$X=\bigl\{ x \in {\Bbb R}^n: \quad |h(x)| \ \leq 1 \ \bigr\} \quad \text{and} \quad \overline{X} ={\Bbb R}^n \setminus X.$$ 
Applying Part 2 of Theorem 6.1 with $F=L^{-1} h$,  $a^2=n$ and $r=L^{-1}$,  we conclude that 
$$\mu\left(\overline{X}\right) \ \leq \ 2 \exp\left\{ -{1 \over 4} \min\left\{{1 \over L},\  {1 \over 4n L^2} \right\} \right\} \ \leq \ 2 e^{-9}. \tag6.2.1$$
Denoting by $[\overline{X}]$ the indicator of $\overline{X}$ and using the H\"older inequality with 
$$p=18 \quad \text{and} \quad q={18 \over 17},$$
we get 
$$ \int_{\overline{X}} e^g \ d \mu = \EE \left([ \overline{X}] e^g \right) \ \leq \ \left(\EE [\overline{X}]\right)^{1/q} \left(\EE e^{pg}\right)^{1/p}=
\left(\mu\left(\overline{X}\right)\right)^{1/q} \left(\EE e^{pg}\right)^{1/p}. \tag6.2.2$$
Applying Part (1) of Theorem 6.1 to $F=L^{-1} g$, $\lambda =pL=18L$ and $a^2=n$, we conclude that 
$$\left(\EE e^{pg}\right)^{1/p} \ \leq \ \exp\left\{ 4n \cdot 18 \cdot {1 \over 12^2 n}  \right\}  =e^{1/2}. \tag6.2.3$$
Combining (6.2.1) - (6.2.3), we conclude that 
$$\int_{\overline{X}} e^g \ d \mu \ < 2 \exp\left\{ - {17 \over 2} +{1 \over 2} \right\} = 2 e^{-8}.  \tag6.2.4$$
On the other hand, for each $x \in X$, the argument of $e^{f(x)}$ does not exceed 1, and hence 
$$\split \left| \int_X e^f \ d \mu \right| \ \geq \ &\Re\thinspace \int_X e^f \ d \mu = \int_X \Re\thinspace e^f \ d \mu \\ \geq \ &(\cos 1) \int_X e^g \ d \mu. 
\endsplit$$
The proof is finished as in Lemma 5.2.
Let 
$$a=\int_X e^g \ d\mu \quad \text{and} \quad b =\int_{\overline{X}} e^g \ d \mu.$$
As in the proof of Lemma 5.2, we have 
$${\left| \EE e^f \right| \over \EE \left| e^f \right|} \ \geq \ (\cos 1) (1-b) -b.$$
Since by (6.2.4), we have $b < 2 e^{-8}$, we conclude that 
$${\left| \EE e^f \right| \over \EE \left| e^f \right|} \ \geq \ (\cos 1) \left(1 - 2e^{-8}\right)- 2e^{-8}  \approx 0.539268878 \ > \ {1 \over 2}.$$
{\hfill \hfill \hfill} \qed 
\enddemo

Now we are ready to prove Theorem 1.3.

 We fix $r$ and prove by induction on the number $m$ of functions the following stronger statement.
\bigskip
{\bf (6.3) Claim:} Let $\phi_1, \ldots, \phi_m: {\Bbb R}^n \longrightarrow {\Bbb C}$ be functions, with respective Lipschitz constants $L_1, \ldots, L_m$, as in Theorem 1.3, and let $\widehat{\phi}_m: {\Bbb R}^n \longrightarrow 
{\Bbb C}$ be yet another function, which is $L_m$-Lipschitz and depends on a subset or all of the coordinates that $\phi_m$ depends on. Suppose further that 
$$\left| \phi_m(x)-\widehat{\phi}_m(x)\right| \ \leq \ \tau \quad \text{for all} \quad x \in {\Bbb R}^n \tag6.3.1$$
and some $\tau >0$. Then we have 
$$\EE \exp\left\{\widehat{\phi}_m + \sum_{i=1}^{m-1} \phi_i \right\} \ne 0, \quad 
\EE \exp\left\{  \phi_m + \sum_{i=1}^{m-1} \phi_i \right\} \ne 0 \tag6.3.2$$
and the ratio of the two numbers (6.3.2) can be written as $e^{\alpha}$ for some $\alpha \in {\Bbb C}$ such that 
$$|\alpha| \ \leq \ 2  \tau. \tag6.3.3$$
\bigskip
{\bf (6.4) Base $m=1$.} We drop the index 1 and denote the functions in question just by $\phi$ and $\widehat{\phi}$ respectively and denote $L_1$ just by $L$. The coordinates that $\phi$ and $\widehat{\phi}$ depend on lie in a subset $\left\{ \xi_j: \ j \in J\right\}$ with $|J| \leq r$, and hence without loss of generality we consider $\phi$ and $\widehat{\phi}$ as functions $\phi, \widehat{\phi}:\ {\Bbb R}^r \longrightarrow {\Bbb C}$. 

We define 
$$\tilde{\phi}_s = (1-s) \phi + s \widehat{\phi} \quad \text{where} \quad 0 \ \leq \ s \ \leq \ 1,$$
so that $\tilde{\phi}_0 = \phi$ and $\tilde{\phi}_1 = \widehat{\phi}$. The functions $\tilde{\phi}_s$ are $L$-Lipschitz. 
Since $L< 1/12 \sqrt{r}$ and $r \geq 9$, by Lemma 6.2, we have 
$$\left| \EE e^{\tilde{\phi}_s} \right| \ \geq \ {1 \over 2} \EE \left| e^{ \tilde{\phi}_s}\right| \quad \text{for all} \quad 0 \leq s \leq 1. \tag6.4.1$$
Then $\EE e^{ \tilde{\phi}_s} \ne 0$ for all $s$ in some neighborhood of the interval $[0, 1] \subset {\Bbb C}$ and we can take a continuous branch of the function $s \longmapsto \ln \EE e^{ \tilde{\phi}_s}$ in that neighborhood.
Then
$$\ln \EE e^{\widehat{\phi}} - \ln \EE e^{\phi} = \int_0^1 \left( {d \over ds} \ln \EE e^{ \tilde{\phi}_s} \right) \ ds =
\int_0^1 {\EE \left( {\widehat{\phi} - \phi} \right) e^{ \tilde{\phi}_s}  \over \EE e^{ \tilde{\phi}_s}} \ ds. \tag6.4.2$$
We have 
$$\left| \EE \left( {\widehat{\phi} - \phi} \right) e^{ \tilde{\phi}_s} \right| \ \leq \ \EE \left| \left( {\widehat{\phi} - \phi} \right) e^{ \tilde{\phi}_s}\right| \ \leq \ \tau \EE \left| e^{\tilde{\phi}_s} \right|.$$
It then follows from (6.4.1) and (6.4.2) that 
$$\left| \ln \EE e^{ \widehat{\phi}} - \ln \EE e^{\phi} \right| \ \leq \ 2  \tau$$
and (6.3.3) follows.
\bigskip
{\bf (6.5) Induction step $m-1 \Longrightarrow m$ for $m \geq 2$.} Let $J \subset \{1, \ldots, n\}$ be the set of indices $j$ such that $\phi_m$ depends on 
$\xi_j$, so that $|J| \leq r$,  and let $\overline{J}=\{1, \ldots, \} \setminus J$ be its complement. 

If $J=\emptyset$ then $\phi_m$ and $\widehat{\phi}_m$ are constants and the ratio of the two expectations (6.3.2) can be written as $e^{\alpha}$ for $\alpha = \pm  \left(\widehat{\phi}_m -\phi\right)$, where the sign is chosen depending on the order in which the ratio is taken. From (6.3.1), we conclude that $|\alpha| \leq \tau$.

If $\overline{J} = \emptyset$, we can consider all functions $\phi_m, \widehat{\phi}_m$ and $\phi_i$, $i=1, \ldots m-1$, as functions ${\Bbb R}^r \longrightarrow {\Bbb C}$, $r \geq 9$. Similarly to Section 6.4, for $0 \leq s \leq 1$, we introduce a function $f_s: {\Bbb R}^r \longrightarrow {\Bbb C}$ by 
$$f_s=(1-s)\phi_m + s \widehat{\phi}_m+ \sum_{i=1}^{m-1} \phi_i.$$
Suppose that $x, y \in {\Bbb R}^r$, $x=\left(\xi_1, \ldots, \xi_r\right)$ and $y=\left(\eta_1, \ldots, \eta_r \right)$ are two vectors that differ in the $k$-th coordinate only, 
so $\xi_j =\eta_j$ for $j \ne k$. Since the function $(1-s) \phi_m + s \widehat{\phi}_m$ is $L_m$-Lipschitz and the functions $\phi_i$, $i=1, \ldots, m-1$ are $L_i$-Lipschitz, we have 
$$\split &|f_s(x)-f_s(y)|  \ \leq \ L_m | \xi_k - \eta_k| +  |\xi_k - \eta_k| \sum\Sb 1 \leq i \leq m-1 \\ i \in I_j \endSb L_i \\
&\qquad \leq \ {1 \over 24 \sqrt{r}} |\xi_k- \eta_k|. \endsplit$$
Hence $f_s$ is ${1 \over 24 \sqrt{r}}$-Lipschitz, $r \geq 9$ and by Lemma 6.2, we have 
$$|\EE e^{f_s} | \ \geq \ {1 \over 2} \EE \left| e^{f_s}\right| > 0. $$
Then arguing as in Section 6.4, we obtain
$$\split \left| \ln \EE e^{f_1} - \ln \EE e^{f_0} \right| =&\left| \int_0^1 \left( {d \over ds} \ln \EE e^{f_s}\right) \ ds \right| =\left| \int_0^1 { \EE \left(\widehat{\phi}_m - \phi_m\right) e^{f_s} \over \EE e^{f_s}} \ ds \right| \\\ \leq \ &  \tau \int_0^1 {\EE \left| e^{f_s} \right| \over \left| \EE e^{f_s}\right|} \ ds \ \leq \ 2 \tau. \endsplit$$

Hence without loss of generality, we assume that $J \subset \{1, \ldots, n\}$ is a proper subset.

As in Section 5, we represent ${\Bbb R}^n$ as the direct sum ${\Bbb R}^n = {\Bbb R}^J \oplus {\Bbb R}^{\overline{J}}$. Let $I \subset \{1, \ldots, m-1\}$ be the set of indices $i$ such that $\phi_i$ shares a coordinate with $\phi_m$, that is, depends on some $\xi_j$ with $j \in J$. For a vector $x \in {\Bbb R}^J$ and a function $\phi_i$ with $i \in I$, we define a function $\phi(\cdot\ | x): {\Bbb R}^{\overline{J}}\longrightarrow {\Bbb C}$ obtained by restricting the coordinates $\xi_j$ with $j \in J$ to those of $x$. Further, we define 
$\Psi: {\Bbb R}^J \longrightarrow {\Bbb C}$ by 
$$\Psi\left(x\right)=\EE_{\overline{J}} \exp\left\{ \sum_{i\in I} \phi_i(\cdot\ | x)  +  \sum\Sb 1 \leq i \leq m-1 \\ i \notin I \endSb \phi_i \right\}, \tag6.5.1$$ 
where the expectation is taken with respect to the measure $\mu$ in ${\Bbb R}^{\overline{J}}$.

We compare values $\Psi(x')$ and $\Psi(x'')$ for $x', x'' \in {\Bbb R}^J$. Suppose that 
  $x'=\left(\xi_j':\ j \in J \right)$ and $x''=\left(\xi_j'':\ j \in J\right)$ differ in the $k$-th coordinate only, that is $\xi_j' = \xi''_j$ for $j \ne k$. Recall that $I_k$ is the set of indices $i$ such that $\phi_i$ depends on the coordinate $\xi_k$. Hence  $I_k \subset I \cup \{m\}$. Switching from $x$ to $y$ affects only 
the functions $\phi_i(\cdot\ | x)$ in (6.5.1)  for $i \in I_k \setminus \{m\}$.
We have 
$$\left| \phi_i(\cdot\ | x') - \phi_i(\cdot\ | x'') \right| \ \leq \ L_i |\xi_k' - \xi_k''| \quad \text{for} \quad i \in I_k$$
Applying the induction hypothesis $|I_k|-1$ times, we conclude that $\Psi(x') \ne 0$, $\Psi(x'') \ne 0$ and that we can write 
$${\Psi(x') \over \Psi(x'')} = e^{\alpha} \quad \text{where} \quad |\alpha| \ \leq \ 2\sum_{i \in I_k \setminus \{m\}} L_i |\xi_k' -\xi_k''|  =  2 |\xi_k'-\xi_k''| \sum_{i \in I_k \setminus \{m\}} L_i.$$
It follows now that we can choose a continuous branch 
$$\psi: {\Bbb R}^J \longrightarrow {\Bbb C}, \quad \psi(x)=\ln \Psi(x), $$
such that 
$$\aligned &\left| \psi(x') - \psi(x'')\right| \ \leq \ 2 |\xi_k' - \xi_k''| \sum_{i \in I_k \setminus \{m\}} L_i \\
&\qquad \text{provided $x'$ and $x''$ differ in the $k$-th coordinate only}. \endaligned \tag6.5.2$$

We have 
$$\aligned &\EE \exp\left\{  \phi_m + \sum_{i=1}^{m-1}  \phi_i \right\} =\EE_J \exp\left\{ \phi_m + \psi\right\} \quad \text{and} \\
&\EE \exp\left\{  \widehat{\phi}_m +  \sum_{i=1}^{m-1} \phi_i \right\} =\EE_J \exp\left\{ \widehat{\phi}_m + \psi\right\},
\endaligned \tag6.5.3$$
where the expectations in the right hand side are taken with respect to the coordinates $\xi_j$ with $j \in J$. We define 
$$\tilde{\phi}_s=(1-s) \phi_m + s \widehat{\phi}_m \quad \text{where} \quad 0 \ \leq \ s \ \leq \ 1,$$
so that $\tilde{\phi}_0 = \phi_m$ and $\tilde{\phi}_1=\widehat{\phi}_m$.  Suppose now that $x', x'' \in {\Bbb R}^J$, $x'=(\xi_j':\ j \in J)$ and $x''=\left(\xi_j'':\ j \in J\right)$
are two vectors that differ in the $k$-th coordinate only. Using (6.5.2) and that $\tilde{\phi}_s$ is $L_m$-Lipschitz, we conclude that 
$$\split &\left| \left(  \tilde{\phi}_m(x') + \psi(x') \right) - \left( \tilde{\phi}_m(x'') + \psi(x'') \right) \right| \\ &\qquad \leq \ 
L_m |\xi_k'-\xi_k''| + 2 |\xi_k' - \xi_k''|  \sum_{i \in I_k \setminus \{m\}} L_i \\
&\qquad \leq \  2|\xi_k'-\xi_k''|  \sum_{i \in I_k} L_i \ \leq \ {1 \over 12 \sqrt{r}} |\xi_k'-\xi_k''|. \endsplit $$
Therefore, $ \tilde{\phi}_s + \psi$ is ${1 \over 12 \sqrt{r}}$-Lipschitz.
Since $r \geq 9$, from Lemma 6.2, we infer
$$\left| \EE_J \exp\left\{ \tilde{\phi}_s + \psi \right\} \right| \ \geq \ {1 \over 2} \EE_J \left| \exp\left\{  \tilde{\phi}_s + \psi \right\} \right| \ >\ 0. \tag6.5.4$$
Then we choose a continuous branch of the function $s \longmapsto \ln \EE_J  \exp\left\{ \tilde{\phi}_s + \psi \right\} $ in that neighborhood of $[0, 1] \subset {\Bbb C}$ and use (6.5.4) to obtain
$$\split &\left| \ln \EE_J \exp\left\{  \widehat{\phi}_m + \psi \right\} - \ln \EE_J \exp\left\{ \phi_m + \psi \right\}\right|  \\ &\qquad=
\left| \int_0^1 \left( {d \over ds} \ln \EE_J \exp\left\{ \tilde{\phi}_s + \psi \right\} \right) \ ds \right|  \\ 
&\qquad =\left| \int_0^1  {\EE_J \left(\widehat{\phi}_m - \phi_m \right) \exp\left\{  \tilde{\phi}_s + \psi \right\} \over
 \EE_J  \exp\left\{ \tilde{\phi}_s + \psi \right\}}  \ ds \right| \\
 &\qquad \leq \ \tau \int_0^1 {\EE_J \left| \exp\left\{ \tilde{\phi}_s + \psi \right\} \right| \over \left| \EE_J \exp\left\{  \tilde{\phi}_s + \psi \right\} \right| }\ ds \\ 
 &\qquad \leq \ 2 \tau.  \endsplit \tag6.5.5$$
 Combining (6.5.3) and (6.5.5), we conclude the proof of Claim 6.3.
{\hfill \hfill \hfill} \qed

\enddocument
\end